\documentclass[preprint,showpacs,floatfix,eqsecnum]{revtex4}

\usepackage{graphicx}
\usepackage{amssymb,amsfonts,amsmath}

\usepackage[dvips]{color}
\usepackage{bm}

\newcommand{\mn}{\mathbf{n}}
\newcommand{\mP}{\mathbf{n}}
\newcommand{\mN}{\mathbf{\nabla}}
\newcommand{\mm}{\mathbf{n}}
\newcommand{\vl}{v_1}

\newcommand{\bsf}[1]{\textsf{\textbf{#1}}}

\newcommand{\Aa}{A}
\newcommand{\Bb}{B}

\begin{document}

\title{Clusters, asters and collective oscillations in
chemotactic colloids}

\author{ Suropriya Saha$^1$, Ramin Golestanian$^2$ and
Sriram Ramaswamy$^{1,3}$}
\affiliation{$^1$ Department of Physics, Indian Institute of Science, Bangalore 560 012, India}
\affiliation{$^2$ Rudolf Peierls Center for Theoretical Physics, University of Oxford, 1 Keble Road, Oxford OX1 3NP, United Kingdom}
\affiliation{$^3$ TIFR Centre for Interdisciplinary Sciences, 21 Brundavan Colony, Osman Sagar Road, Narsingi, Hyderabad 500 075, India}


\begin{abstract}
The creation of synthetic systems that emulate the defining properties of living
matter, such as motility, gradient-sensing, signalling and replication, is a grand
challenge of biomimetics. Such imitations of life crucially contain active
components that transform chemical energy into directed motion. These artificial
realizations of motility point in the direction of a new paradigm in engineering,
through the design of emergent behavior by manipulating properties at the scale
of the individual components. Catalytic colloidal swimmers are a particularly
promising example of such systems. Here we present a comprehensive theoretical
description of gradient-sensing of an individual swimmer, leading controllably
to chemotactic or anti-chemotactic behavior, and use it to construct a framework
for studying their collective behavior. We find that both the positional and the
orientational degrees of freedom of the active colloids can exhibit
condensation, signalling formation of clusters and asters. The kinetics of
catalysis introduces a natural control parameter for the range of the
interaction mediated by the diffusing chemical species. For various regimes in
parameter space in the long-ranged limit our system displays precise
analogs to gravitational collapse, plasma oscillations and electrostatic
screening.  We present prescriptions for how to tune the surface properties
of the colloids during fabrication to achieve each type of behavior.
\end{abstract}

\maketitle

\section{Introduction}
Dynamic self-organization of motile components can be observed in
a wide range of length scales, from bird flocks \cite{Cavagna2014} to bacterial colonies \cite{Levine2000,Zhang2010} and
assemblies of motor and structural proteins
\cite{Huber2013,Schweitzer2007}. The fascination with these
phenomena has naturally inspired researchers to use a physical understanding of motility
to engineer complex emergent behaviors in model systems that promise revolutionary
advance in technological applications if combined with other novel biomimetic
functions, such as signal processing and decision making \cite{VRobots}, or replication \cite{chaikin_replication}.

Symmetry-based phenomenological theories, coarse-grained or particle-based
\cite{TonerTu1995,SriramReview,MCMetal2013,Chate2004,Vicsek2012,laulub2009,
Bialke2013}
offer a guide to the rich possibilities immanent in self-driven
systems, but designing a system requires a bottom-up approach. Biological
components pose inevitable limitations on this
task, while chemical \cite{shashi2011}, mechanical \cite{Sano_single_grain}
or externally actuated \cite{Dreyfus2005} imitations appear more promising.
In addition to motility, living organisms have developed mechanisms that
allow them to orient their motion in response to chemical gradients, and send
signals to recruit or repel others \cite{Huber2013,KS1971}. Can inanimate matter
imitate these more complex functions? We show that it can, and
present the necessary design principles. We consider the case of catalytic
active colloids \cite{paxton2004,gla2005,Howse2007,Jiang2010}, which we now
describe in brief. Recall that a colloidal particle can be
driven ``phoretically'' into motion by {\em externally imposed} chemical,
electrostatic
or thermal gradients \cite{Anderson1989}. An
{\em active} colloid -- a particle
coated asymmetrically with catalyst and immersed in a {\em
uniform} background of substrate\footnote{For consistency with the
nomenclature
of enzyme catalysis literature we refer to the reactant as `substrate', not to
be confused with other uses of the term.} -- generates its own chemical
gradient \cite{paxton2004,gla2005,Howse2007,Jiang2010,ruckner2007,gla2007}, and
thus moves autonomously in a direction determined by the polarity of
the coat\footnote{A related idea has been investigated in the
context of an ionic cell-motility mechanism
\cite{Mitchell-1,Mitchell-2,lammert}.}.
Such self-phoretic particles, whose individual activity and interactions
one can design, offer the opportunity to create systems with controllable,
emergent collective behavior
\cite{Hong2007,Ibele2009,rg2009,Ibele2010,Kagan2011,Theurkauff2012,
NYU2013,Duan2013,Baraban2013}. To this end it is essential to construct a
description at a coarse-grained level, with coefficients expressed in terms of
single-particle parameters\footnote{When this work was being prepared for
submission, we learned of unpublished results from the groups
of H Stark and J Brady on self-phoretic swimmers interacting through their
diffusion fields. The construction of chemotactic behavior from the patterning
of the colloid, the role of enzyme kinetics, the dynamics of orientation fields
and aster formation, the occurrence of underdamped modes and the possibility of
spontaneous oscillation are among the distinguishing features of our work.}.

Our focus is on how the {center of mass and} orientation vector of an
active colloid {are affected by}
an externally imposed gradient of substrate molecules. Depending on details
of geometry, activity, and mobility \cite{gla2007}, an active colloid will
respond to the local gradient of the substrate concentration through four
distinct mechanisms. (i) {\it Chemotaxis:} The fluid flows set up around the
particle can turn its axis of orientation to align parallel or antiparallel to
the local gradient; this process has active contributions arising from
the chemical reaction as well as passive ones
(ii) {\it Polar run-and-tumble motion:} The enzymatic rate depends nonlinearly
on the local concentration of the substrate with a characteristic Michaelis-Menten
form inherited from the underlying catalytic kinetics of the reactions \cite{Ebbens2012}.
The combination of enhanced activity at high concentrations and randomized orientation
acts to effectively populate the colloids in ``slow'' regions \cite{Cates2012}.
(iii) {\it Apolar run-and-tumble motion:} An active colloid can also chemotax by
a net motion of its center along a gradient in a noise-averaged sense.
(iv) {\it Phoretic response:} The colloid moves along an external
{chemical
gradient by diffusiophoresis.} A summary of the different modes is depicted in
Fig. 1.

Catalytic colloids consume a {substrate and generate} product molecules,
and hence act as mobile sources and sinks of these chemicals in the solution
making their concentration profile nonuniform. In a suspension of such active
colloids, each individual responds---via the above four mechanisms---to
the gradients produced by other colloids due to their activities.
The various contributions are independent of each other and their balance will
be modified as we move in the space of control parameters, leading to a variety
of collective behaviors. In particular, we highlight the intriguing
possibility that the positional and orientational degrees of freedom could
exhibit different and independent types of order depending on the parameters,
as shown in Figs. 2 and 3.

\begin{figure}[t]
\begin{center}
\includegraphics[angle=0,width=.8 \columnwidth]{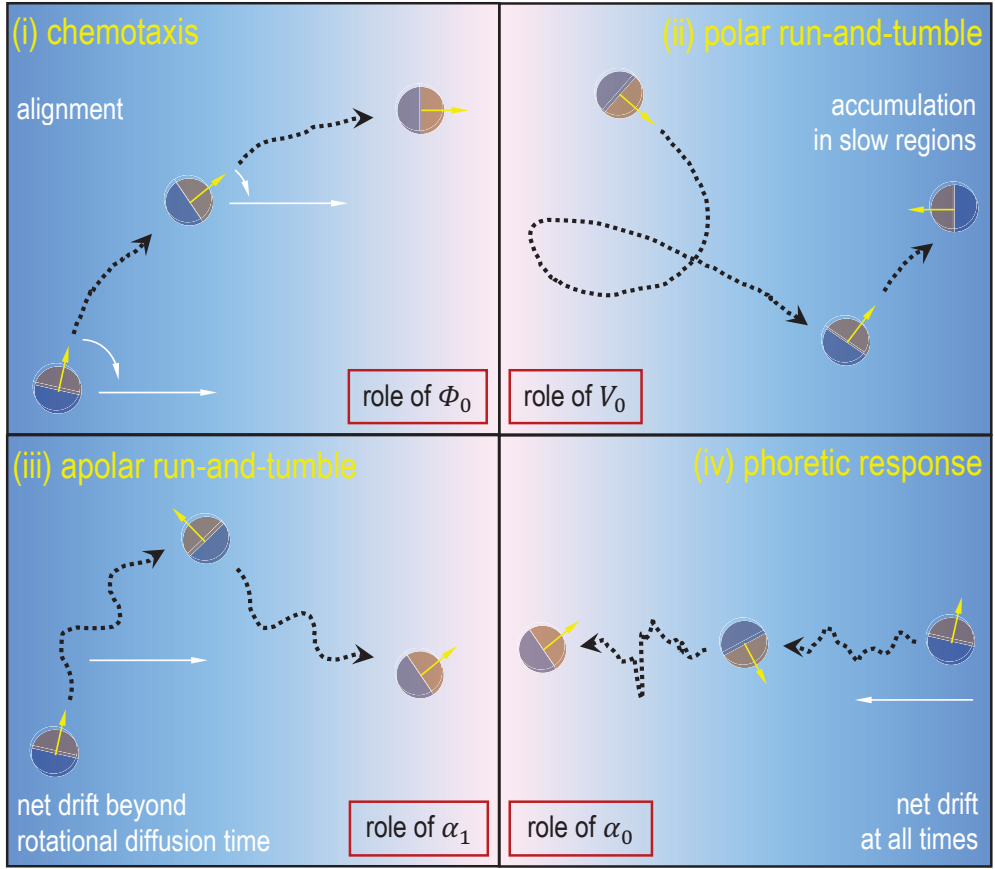}
\caption{A {schematic} summary of the four different ways a single swimmer
responds
to gradients corresponding to the different terms in Eq. \ref{res_sp}. In each
panel, three consecutive snapshots (with equal time intervals) are sketched
together {with typical connecting trajectories.}
In (i) and (ii) the polarity of the colloid controls the direction of
motion. In (iii) the motion will be along the main symmetry axis of the colloid but driven by the gradient (hence the colloid can move forward
or backward instantaneously). In (iv) the motion is independent of the
polarity and symmetry axis of the colloid. Processes (i) and (iv)
represent steady angular and linear drift while {in} (ii) and (iii) the
{gradient-seeking} behavior is assisted by noise. Each mechanism is
controlled by the relevant spherical harmonics coefficient of the
{material-dependent} particle mobility {and activity}, which can be
modified by
construction.}
\end{center}
\label{fig:scheme}
\end{figure}

We consider a fluid medium containing a concentration $s(\mathbf{r},t)$ of substrate (S)
molecules, which upon contact with a catalyst are converted to a product P with
concentration $p(\mathbf{r},t)$. The rate of conversion $\kappa$ obeys
Michaelis-Menten kinetics \cite{MM}, growing linearly at small $s$ and crossing
over to saturation for sufficiently large values of $s$. Our main results are as
follows: {\bf (1)} for the case of a uniform gradient of substrate, we establish the form
of the angular velocity $\bm{\omega}$ induced on a single catalytic colloid as
a function of the spherical harmonic components of the activity $\sigma$
and the mobilities $\mu_{s}$ and $\mu_{p}$ corresponding to S and P.
We can therefore propose criteria for the surface patterning required to produce
chemotactic and anti-chemotactic motion. We also find the various contributions to
the translational velocity $\bm{v}$ of the colloid, arising from self-propulsion
and drift due to the external gradient. Explicitly, we find
\begin{eqnarray} \label{res_sp}
\bm{\omega}&=&\Phi_0(\sigma,\mu_p,\mu_s) \; \hat{\mn} \times \bm{\nabla} s, \\ \nonumber
\mathbf{v}&=& V_0(s) \hat{\mn} - \alpha_0 \bm{\nabla} s - \alpha_1 \hat{\mn} \hat{\mn} \cdot \bm{\nabla} s.
\end{eqnarray}
where the definitions of the coefficients in terms of the surface properties of the colloids
are given below. {\bf (2)} We use our results for a single particle in an external gradient
to construct the collective equations of motion for the number density and orientation of
the colloids in a uniform medium, interacting via their effect on the substrate and product
concentration fields. The interplay between self-propulsion, phoretic drift, and alignment,
driven by and mediated via chemicals, falls into two distinct regimes. (a) When the fuel
concentration is small enough such that the catalytic activity is {\it diffusion-limited},
the chemical concentrations will be effectively screened, and the system could develop
enhanced number fluctuations and clumping instabilities (where all wavelengths above
a threshold are unstable) and patterns with a given length scale (where the fastest-growing
mode has finite wavelength). (b) At sufficiently high fuel concentrations where the catalytic
activity becomes {\it reaction-limited}, the chemical fields are not screened and can mediate
long-ranged interactions that could lead to a wider variety of instabilities. In particular,
for the case of effectively attractive phoretic interaction we observe collapse transitions
that are dissipative analogs of a Jeans instability \cite{JeansInstability}, with or without
simultaneous condensation of asters. For effectively repulsive phoretic interactions, we observe
counter-intuitively that a collapsed phase with aster condensation is still possible, as are
stable phases exhibiting Debye-like screening, similar to electrolytes. Moreover, we find that
in this regime the system could exhibit plasma-like oscillations in response to perturbations,
or spontaneous, self-sustained ringing. Detailed phase diagrams inferred from our stability
analysis and structure factor calculations are found in Figs. 2 and 3. These
are parameterized, via definitions in Eq. \ref{ab} below, by coefficients $A$ 
describing the chemotactic response from panels (i) and (ii) of
Fig. 1, and $B$ the phoretic response from panels (iii) and (iv) of that
figure. 
We now show how we obtained these results.

\subsection{{Background: diffusiophoresis and self-diffusiophoresis}}
\label{phoresis}
Diffusiophoresis\footnote{We will not discuss similar phoretic propulsion
mechanisms through gradients in temperature or electrostatic potentials.} is the
force-free, torque-free propulsion of a colloid by a solute
concentration gradient \cite{Anderson1989}. In a fluid of viscosity $\eta$ at
temperature $T$ a species with concentration $c$ interacting through an
effective potential $\Psi$ with a particle surface with normal along the local
$z$ axis gives rise, via the Stokes equation, to a surface ``slip velocity''
$\mathbf{v}_{\rm slip} =  \mu {\mathbf{\nabla}}_{\parallel}c$, with the phoretic
mobility $\mu = \frac{k_B T}{\eta} \int_{0}^{\infty} z (1-e^{-{\Psi}/{k_B T}})
dz$ that can have either {\em sign} depending on  $\Psi$ \cite{FairAnd1989}.
We are also interested here in \textit{self}-diffusiophoresis \cite{gla2005}
that occurs when $\nabla c$ is not imposed externally but generated by processes
on the particle itself.

\begin{figure}[t]
\begin{center}
\includegraphics[angle=0,width=.8 \columnwidth]{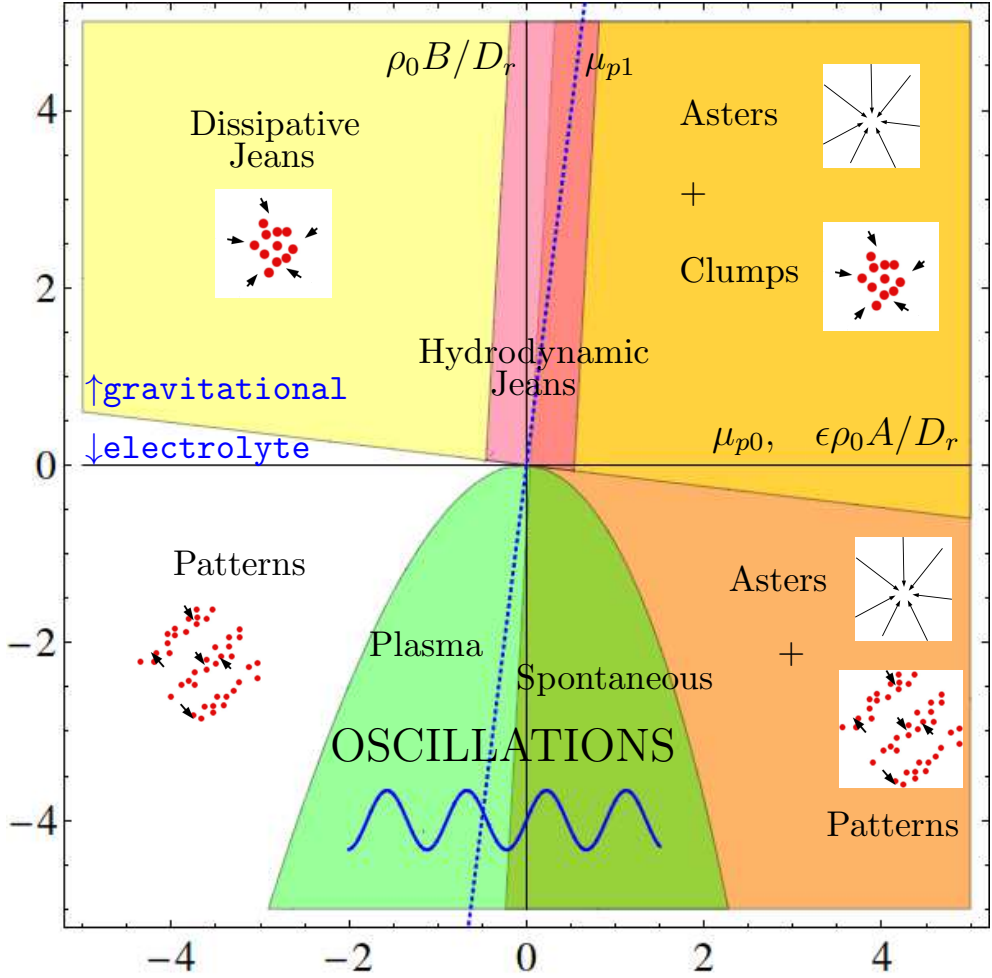}
\caption{The phase diagram in the {\em reaction-limited} regime (with abundant fuel) shows
a variety of possible states in the parameter space spanned by suitably non-dimensionalized effective
chemotactic ($A$) and phoretic ($B$) response coefficients, defined in Eq. \ref{ab}.
The dashed line and the $A$-axes correspond to independent changes of the
fundamental and first harmonic ($\mu_{p0}$ and $\mu_{p1}$) of the mobility
corresponding to the reaction product, respectively. They represent possible
experimental paths that can be explored in sequences of experiments on particles
designed with suitable mobility coats.}
\end{center}\label{phasedia1}
\end{figure}

\subsection{A single chemotactic motile colloid} \label{singlechemo}
Consider a single swimmer, whose mobilities and catalytic coat have the
common symmetry axis $\hat{\bf n}$.
When placed in a \textit{uniform} substrate background, such a particle moves in a direction
determined by $\hat{\bf n}$ if the coatings are sufficiently asymmetric \cite{gla2007}. What happens
in an inhomogeneous {background?}
Can the flows set up by the interaction of S and P with the swimmer
surface {reorient its} axis $\hat{\bf n}$ with respect to
the local concentration gradient, {thus imitating} chemotaxis? To answer
this question,
we solve for the concentrations $s$ and $p$, {with diffusivities $D_s$ and
$D_p$ respectively}. We incorporate the catalytic chemical reaction S $\to$ P
through source and sink boundary conditions on particle fluxes normal ($\perp$) to the swimmer surface:
\begin{eqnarray}
-D_s \mathbf{\nabla}_{\perp} s = -\kappa_1 s  P_s \sigma(\theta,\phi); \,\,\,\, -D_p \mathbf{\nabla}_{\perp} p = \kappa_2 P_p \sigma(\theta,\phi),
\label{bc}
\end{eqnarray}
{where $P_p(\theta,\phi) \equiv 1-P_s(\theta,\phi)$ is the probability that
the
enzyme at $(\theta, \phi)$ is bound to the substrate.} Stationarity implies
$\kappa_1 s  P_s = \kappa_2 P_p$ leading to the Michaelis-Menten \cite{MM}
expression
\begin{math} 
\kappa_2 P_p \equiv \kappa(s) \equiv {\kappa_2 \kappa_1
s}/{(\kappa_2 + \kappa_1 s)}
\end{math}
for the reaction velocity per molecule.
Number conservation for the products and substrates, and the assumption that $s$
and $p$ diffuse rapidly compared to the colloid so that time dependencies
and advection by flow \cite{gareth} can be ignored give $D_p p+D_s s=D_s s_b$,
where $s_b$ is the {background} substrate profile. We thus need to solve
for just
one of the two concentration fields.
{We work in the linear
regime\footnote{Results
from another limit of interest, $D_p \gg D_s$ and $s_b \ll
\kappa_1/\kappa_2$,
in which the chemical reaction influences significantly the local value of $s$
so that the coupled dynamics of orientation and translation can lead to
oscillations, will be discussed elsewhere.} $s_b \ll \kappa_1/\kappa_2$, where
the profile of product $p$ resulting from this process is sensitive
to the imposed gradient of $s$, and in the limit
where S diffuses rapidly so that its profile is maintained.} The resulting slip
velocity, which
has contributions from both the substrate and the product, leads to the linear
and angular velocities
\begin{math}
\bm{\omega} = -\frac{3}{16 \pi R } \int  \hat{\mathbf{r}}
\times
\mathbf{v}_{\rm slip} (\mathbf{r}) \; \mbox{d} \Omega
\end{math}
and
\begin{math}
\mathbf{v} = -\frac{1}{4 \pi} \int \mathbf{v}_{\rm slip} (\mathbf{r}) \; \mbox{d}
\Omega
\end{math}
for spherical colloids.

To understand the general trends in chemotactic behavior arising
from simple catalytic patterns, we work with a limited number of non-zero
spherical harmonic components of $\sigma$ and $\mu_p$. For example, taking
$\sigma_l, \mu_{pl}=0$ for $l\geq 3$ we find the expression for the angular velocity
given in Eq. \ref{res_sp} with
\begin{equation} \label{spFom}
\Phi_0 = -\frac{3 \mu_{s1}}{4 R}-\frac{\kappa_1}{60 D_p} \; \left(5 \mu _{p1}
\sigma _0 + 2 \mu _{p2} \sigma _1-\mu _{p1} \sigma _2 \right),
\end{equation}
where a negative (positive) value corresponds to chemotactic
(antichemotactic) response. 
The first term on the left is the passive response to the external gradient 
due to a polarity in $\mu_s$ alone while the second term is the active contribution 
involving both $\sigma$ and $\mu_p$. The form of Eq. \ref{spFom} serves to
illustrate some features that hold even without the truncated expansion in
$l$: if either $\sigma$ or $\mu_p$ contain all odd or all even harmonics
there is no reorientation in response to the gradient, a result which holds
for spheroidal swimmers as well. The expression for the product contribution
in Eq. \ref{spFom} is a sum of products of $\sigma_l$ and $\mu_{p,l\pm1}$,
which can be used to design chemotactic colloids with a desired response.
Lastly, regardless of the form of $\sigma$, $\omega = 0$ for $\mu_p$ uniform
over the sphere.

\begin{figure}[t]
\begin{center}
\includegraphics[angle=0,width=.9 \columnwidth]{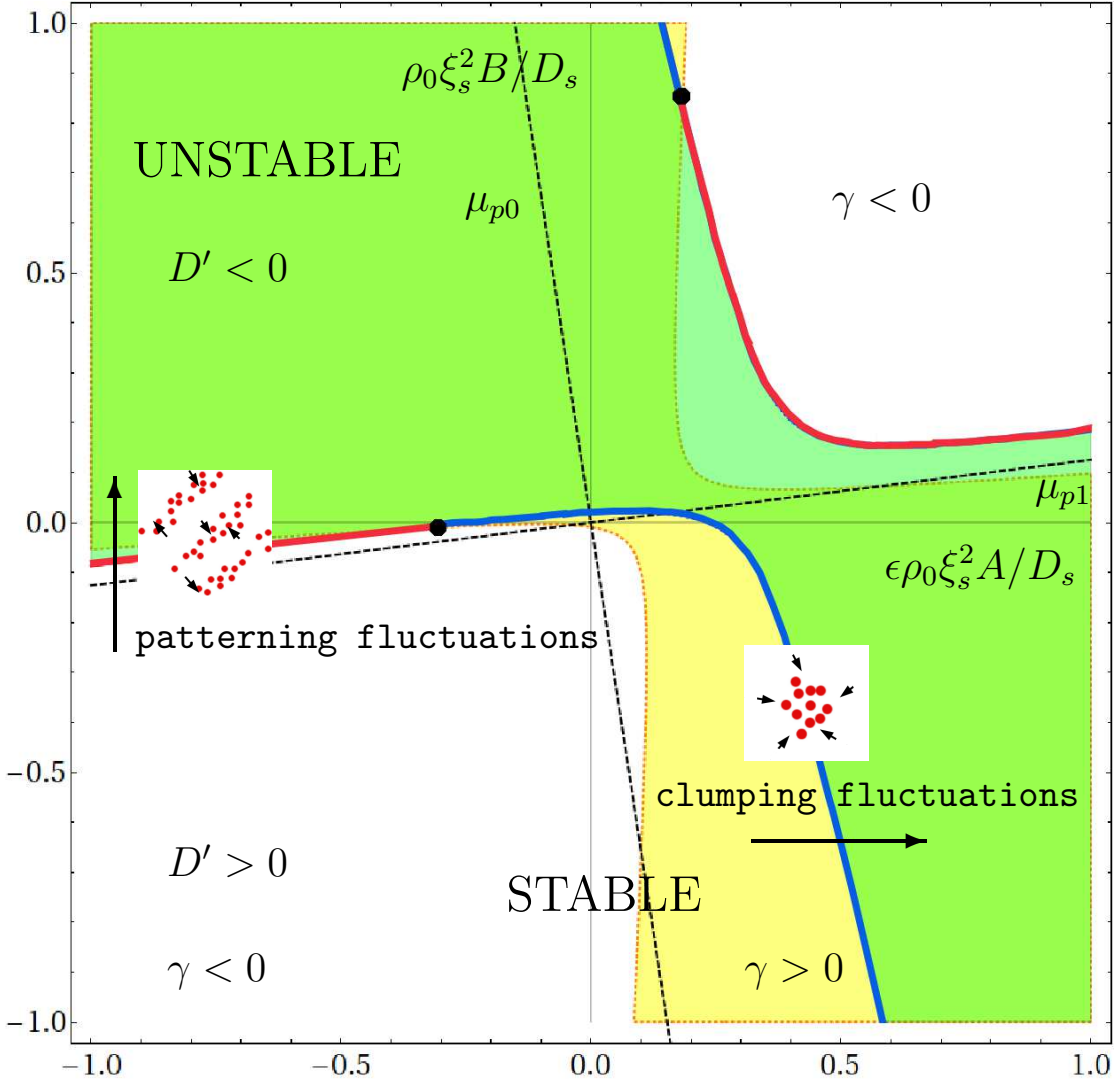}
\caption{The {\em diffusion-limited} regime (with limited fuel supply)
exhibits different types of instability in the parameter space spanned by suitably 
non-dimensionalized effective
chemotactic ($A$) and phoretic ($B$) response coefficients, defined in Eq. \ref{ab}.
The effective diffusivity $D'$ (defined in Eq. \ref{moseSc}) is negative in the green region
and positive in the white or yellow regions, signalling the presence of an instability upon
going from the yellow or white to green. The dominant fluctuations on
approaching the instability signal a tendency to form modulations 
(across the red line) or clumps (across the blue line) state, depending on the
sign of the parameter $\gamma$ (defined in Eq. \ref{Srho}).
The character of the fluctuations on the stability boundary changes at
the location shown by the black dot.
The axes corresponding to possible experimental changes of the fundamental and first harmonic ($\mu_{p0}$ and $\mu_{p1}$) of the mobility
corresponding to the reaction product, respectively, are shown by two dashed lines.}
\end{center}\label{phasedia2}
\end{figure}

We also obtain the net translational velocity $\mathbf{v}$, as in Eq. \ref{res_sp}
where
\begin{eqnarray}\label{spFlin}
&& \hskip-0.6cm V_0 = {\frac{\kappa_1 s_b}{15 D_p}}(5 \sigma_1 \mu_{p0}+ 2
\sigma_2 \mu_{p1}
-\sigma_1 \mu_{p2}),  \nonumber \\
&& \hskip-0.6cm \alpha_0 = - (\mu_{s0} + \frac{1}{10} \mu_{s2}) - \frac{ \kappa_1 R }{10 D_p}
(\sigma_0 \mu_{p2}-{2 \over 9} \sigma_1 \mu_{p1}-2 \sigma_2
\mu_{p0}\nonumber \\
&&+{1 \over 35} \sigma_2 \mu_{p2}), \nonumber \\
&& \hskip-0.6cm \alpha_1 = - \frac{1}{10} \mu_{s2} -
\frac{\kappa_1 R}{ 30 D_p} (10 \sigma_0 \mu_{p0}+ \sigma_0 \mu_{p2}+2
\sigma_1 \mu_{p1}-2 \sigma_2 \mu_{p0}   \nonumber \\
&& +{29 \over 35} \sigma_2 \mu_{p2}).
\end{eqnarray}
The three contributions to the translational velocity correspond to self-propulsion
(along $\mathbf{\hat{n}}$), phoretic drift (along $\mathbf{\nabla} s$), and an anisotropic drift
that is instantaneously along $\mathbf{\hat{n}}$, but leads to net motion along $\mathbf{\nabla} s$
as rotational noise de-correlates $\mathbf{\hat{n}}$. The latter amounts to a contribution
to run-and-tumble gradient-seeking motion, which we name apolar run-and-tumble; see Fig. 1.

To demonstrate how the chemotactic response of catalytic colloids can be designed,
we have calculated $\Phi_0$ for an example of swimmers with uniform spheroidal caps
of catalytic and mobility patterns as $\sigma(\theta) \propto \Theta(\theta - \theta_1)$ and
$\mu_p \propto 1+\Theta(\theta - \theta_2)$. Figure 4 shows $\Phi_0$
as a function of $\theta_1$ for different values of $\theta_2$. For $\theta_2 = \pi/2$,
$\Phi_0$ is antisymmetric as a function of $\theta_1$. For a given $\theta_2$, $\Phi_0$
peaks near $\theta_1=\theta_2$ as the slip velocity is maximum when the position where
$\mu_p$ is maximum coincides with the region where $p$ changes most rapidly, which for
the given form of $\sigma$ is at $\theta_1$. This example showcases the possibility to control
the response of individual catalytic colloids, and thus their collective behaviors, by
following design rules that include varying systematically their geometric features.
We now combine the individual responses of active colloids to construct a theoretical
description for their collective behaviors.

\section{From chemotaxis to collective motion} \label{collective}
Catalytic swimmers of the type discussed above interact through the S and P
chemical fields as well as via hydrodynamics
\cite{Enkeleida2012,Stark2013,ELauga}. We restrict
our attention to their chemotactic interaction, and construct the collective
behavior of many swimmers by looking at pairwise interactions. Consider,
therefore, two swimmers separated by a distance
$r$ in a uniform medium of substrate molecules. The {reaction} S $\to$ P
that takes place on the surface of each swimmer modifies the $s$ field
as seen by the others, and each is also a source for P. In the absence of a
background of other swimmers, inhomogeneities in the $s$ and $p$ fields in
steady state decay as $1/r$. Each swimmer senses and responds to the
magnitude and the gradient of $s$ through the motility and chemotaxis
mechanisms outlined above. In addition, each particle responds to the $p$
field produced by the reactions on the surfaces of all the particles,
just as it would to any externally imposed solute gradient \cite{Anderson1989}.
The resulting equations of motion for the position $\mathbf{r}_{\alpha}$
and orientation unit vector $\mathbf{\hat{n}}_{\alpha}$ of the $\alpha$th
{, to linear order in $\nabla s$, $\nabla p$,} take the form
\begin{eqnarray}
\frac{\mbox{d} \mathbf{r}_\alpha}{\mbox{d} t} &=& V_0(s)
\mathbf{\hat{n}}_\alpha - \alpha_0 \mathbf{\nabla}s
- \alpha_1 \mathbf{\hat{n}}_\alpha
\mathbf{\hat{n}}_\alpha \cdot \mathbf{\nabla}s +
\beta_0 \nabla p \nonumber \\
&& + \beta_1 \mathbf{\hat{n}}_\alpha \mathbf{\hat{n}}_\alpha \cdot
\mathbf{\nabla} p + \sqrt{2D} \; \mathbf{f}^{r}_\alpha(t),   \nonumber \\
 \frac{\mbox{d} \mathbf{n}_\alpha}{\mbox{d} t} &=& \Phi_0
(\mathbf{\hat{n}}_\alpha \times
\mathbf{\nabla}s)\times \mathbf{\hat{n}}_\alpha + \Omega_0
(\mathbf{\hat{n}}_\alpha \times \mathbf{\nabla}
p)\times \mathbf{\hat{n}}_\alpha \nonumber \\
&& + \sqrt{2D_r} \; \mathbf{\hat{n}}_\alpha
\times\mathbf{f}^{n}_\alpha(t),
\label{polposeq}
\end{eqnarray}
where additional coupling constants
\begin{equation}\label{spFGradp}
\Omega_0 = -\frac{3 \mu_{p1}}{4 R},  \;\; \beta_0 = -(\mu_{p0} + \frac{1}{10}\mu_{p2}),
\;\; \beta_1 =- \frac{1}{10} \mu_{p2},
\end{equation}
are introduced to take account of the response of each colloid to a product gradient
produced by the others. {In Eq. \ref{polposeq} thermal as well as active
fluctuations are included phenomenologically via Gaussian unit-strength white
noise terms $\mathbf{f}^r_\alpha, \, \mathbf{f}^n_\alpha$, with strengths $D$
and $D_r$.} $\Phi_0 >0$ and $\Omega_0>0$ correspond to
swimmers that respond chemotactically to $\mathbf{\nabla}s$ and
$\mathbf{\nabla}p$ {respectively}. $\alpha_0 > 0$ and $\beta_0>0$ imply
attractive contributions to the interactions between the swimmers due to $s$ and
$p$ {respectively}. $V_0(s)>0$ by definition as we choose
$\hat{\mathbf{n}}_{\alpha}$ to point in the direction
in which a solitary swimmer moves. The form of Eq. \ref{polposeq} follows on
general grounds of symmetry. The point of our calculation is that it gives
explicit expressions for the tactic and phoretic mobilities and the expressions for
$\alpha_i$ and $\beta_i$. Moreover, the substrate and product fields are themselves determined by the
distribution of colloid positions and orientations.
The substrate is consumed and the product is generated at the rate
\begin{math}
\label{rates}
Q(\mathbf{r},t) = \kappa(s)
\sum_{\alpha} \int_{|\mathbf{X}_{\alpha}| = R}
\delta(\mathbf{r}-\mathbf{r}_\alpha-\mathbf{X}_{\alpha}) \sigma
(\mathbf{X}_{\alpha} \cdot \mathbf{\hat{n}}_\alpha),
\end{math}
where $\mathbf{X}_{\alpha}$ is the position coordinate on the $\alpha$th swimming
sphere of radius $R$, and the catalytic coat $\sigma$ is expressed in lab-frame
coordinates. We assume the system is maintained\footnote{We assume prompt
replenishment of consumed substrate, and we work on timescales long enough that
the product has reached the sample boundaries, where it is absorbed.} in a
steady state with mean substrate concentration $s_0$ and develop $Q$ to leading
orders in a gradient expansion to obtain a coarse-grained description. We begin by
relating the $s$ and $p$ fields to the coarse-grained density and orientation fields
of the colloids, namely,
$\sum_\alpha \delta(\mathbf{r} - \mathbf{r}_\alpha) = \rho(\mathbf{r})$ and
$\sum_\alpha \mathbf{\hat{n}}_\alpha \delta(\mathbf{r} - \mathbf{r}_\alpha)
 = \mathbf{w}(\mathbf{r})$ (see the Supplementary Material). We find
\begin{eqnarray}
(\partial_t  - D_s \nabla^2) s = -N \kappa(s) (\rho -\epsilon \mathbf{\nabla}\cdot \mathbf{w})= -(\partial_t  - D_p \nabla^2) p.
\label{psscale}
 \end{eqnarray}
In Eq. \ref{psscale}, $N=4 \pi R^2 \sigma_0$ is the total number of
enzymatic sites on the surface of the swimmer and $\epsilon = R \sigma_1/3
\sigma_0$ measures the degree of polarity of the catalytic coat. 
We work in the limit of $\partial_t s=\partial_t p=0$. Linearizing Eq. \ref{psscale}
around a steady state with ($\rho_0, s_0, p_0$),
we find the following results for the Fourier components of the concentration at wavevector $\mathbf{q}$:
 \begin{math}
s_{\mathbf{q}} = -{N \kappa(s_0)(\rho_{\mathbf{q}} -\epsilon i \mathbf{q} \cdot \mathbf{w}_{\mathbf{q}})}/{[D_s (q^2 +
\xi_s^{-2})]}, \;
   p_{\mathbf{q}} = -{D_s s_\mathbf{q}}/{D_p},
\label{SPsol}
  \end{math}
where we introduce {\it the screening length}
\begin{equation}
\label{xis}
\xi_s = [N \rho_0 \kappa'(s_0)/D_s]^{-1/2},
\end{equation}
that is a measure of the range of interactions mediated by S and P.
\begin{figure}[t]
\begin{center}
\includegraphics[angle=0,width=.5 \columnwidth]{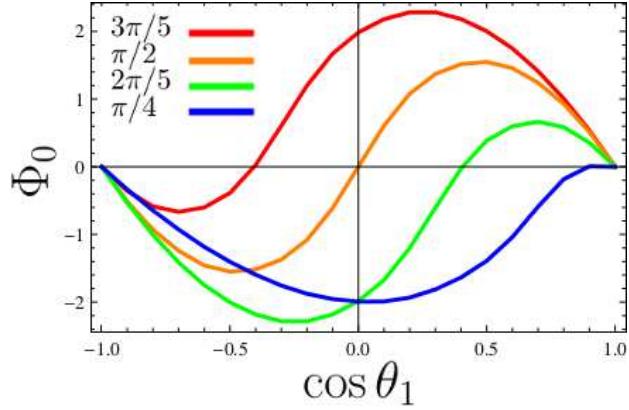}
\caption{Angular velocity coefficient for spherical colloids as a function of $\theta_1$ (that parameterizes
the size of the catalytic coating) for different values of $\theta_2$ (that parameterizes
the mobility pattern) quoted in the legend.}
\end{center}                       \label{omega_shape}
\end{figure}
For $s_0 \ll \kappa_2/\kappa_1$, i.e. on the linear or \textit{unsaturated}
part of the MM curve, $\xi_s$ is finite and the interactions are
therefore short-ranged. For $s_0 \gg \kappa_2/\kappa_1$, i.e. on the
\textit{saturated} part of the MM curve, $\xi_s \to \infty$ and the product
mediates an effective long-ranged interaction amongst the colloids.

{Starting from the Langevin equations in Eq. \ref{polposeq}, we next
construct
equations of motion for $\rho$ and $\mathbf{n}(\mathbf{r}) = \mathbf{w}/\rho$ 
as defined above Eq. \ref{psscale}. 
While coarse-graining, we see inevitably that the dynamical equation for $\mathbf{n}$
involves higher moments of the orientational distribution function, which must be
re-expressed in terms of lower moments \cite{bertin2006,shelley} using an appropriate 
closure which we discuss in the Supplementary Material. 
The S and P fields produced by inhomogeneities in density and the 
divergence of the polar order parameter mediate interactions between
swimmers through
$\rho$ and the longitudinal component $\mP_{Lq}=\hat{\mathbf{q}}\hat{\mathbf{q}}
\cdot \mP_{q}$. The linearized dynamics in the isotropic phase closes in terms
$\rho$ and $\mn_{L \mathbf{q}}$, whose coarse-grained equations we present
in the saturated limit and for wavenumbers $q \ll \xi_s^{-1}$ in the
unsaturated case.
We will see that despite the presence of a self-propelled velocity field and a density, 
there is an important contrast relative to models such as Toner-Tu \cite{TonerTu1995}: 
the interactions considered here offer no mechanism to promote flocking, 
i.e., the global parallel alignment of $\mathbf{n}$.}

We have calculated the mode structure and steady state structure factors
$S_\rho = \int_\omega \langle|\rho_{\mathbf{q} \omega}|^2 \rangle $ and $S_n =
\int_\omega \langle |\mP_{L\mathbf{q} \omega}|^2 \rangle$ for both the
unsaturated and the saturated cases by adding {phenomenological gaussian
white noise terms,} conserving for $\rho$ and nonconserving
for $\mP_L$, to the equations of motion. 
The equations of motion and their stability analysis which we now present 
are best shown in terms of 
coefficients
\begin{eqnarray}
&& \hskip-1.7cm A = N \kappa(s_0) \left[\frac{\Omega_0}{D_p} -
\frac{\Phi_0}{D_s} + \frac{V_0(s_0)}{2 D_s} \frac{d \ln \kappa}{d
s}|_{s_0}\right], \nonumber \\
&& \hskip-1.7cm B = N \kappa(s_0) \left[\frac{1}{D_p} (\beta_0 +
\frac{\beta_1}{3}) + \frac{1}{D_s} (\alpha_0 + \frac{\alpha_1}{3})\right],
\label{ab}
\end{eqnarray}
which give the effective chemotactic and phoretic response to gradients
respectively.

\section{Unsaturated}
In the unsaturated case $\xi_s < \infty$, for wavenumbers $q \ll \xi_s^{-1}$, coarse grained equations read
\begin{subequations}
\begin{eqnarray}
&& \hskip-0.7cm [\partial_t +2D_r -(D+\frac{v_1^2 s_0^2}{30 D_r}) \nabla^2
+ (\epsilon \rho_0 \xi_s^2 A - \frac{v_1^2 s_0^2}{90 D_r}) \mN \mN \cdot] \mP_{L} \nonumber \\
&& + \frac{(v_1 s_0 - \rho_0 \xi_s^2 A)}{3 \rho_0} \mN \rho = 0,\label{unsaturateda}\\
&& \hskip-0.7cm [\partial_t- (D-\rho_0 \xi_s^2 B)\nabla^2]\rho_0^{-1} \delta \rho
+ (v_1 s_0- \epsilon \rho_0 \xi_s^2 B \nabla^2) \mN \cdot \mP_L=0, \nonumber \\   \label{unsaturatedb}
\end{eqnarray}
\label{unsaturated}
\end{subequations}
\noindent
where we use $V_0(s_0) \equiv v_1 s_0$ as the self-phoretic velocity scales linearly
with the substrate concentration.

{Several features of Eq. \ref{unsaturated} are noteworthy. From Eq.
\ref{psscale}, P is abundantly available or S is depleted where the
density $\rho$ is high. Phoretic movement up (down) the gradient of $p\,(s)$
can thus lead to a propensity of swimmers to swim up their concentration
gradients and hence a change in the sign of the diffusivity in Eq. \ref{unsaturatedb}
through $\rho_0 \xi_s^2 B$. Since the swimmer preferentially moves along its polar axis, 
$\mP$ can be viewed as a velocity field and the $\mN \rho$ term as a pressure
gradient. 
We see that for large enough $\epsilon \rho_0 \xi_s^2 A$, the signs of coefficients
that are analogous to bulk viscosity, and squared sound speed can change signalling 
an instability and possible novel condensation phenomena whose nature 
will be revealed only by a nonlinear treatment with appropriate noise terms. 
Note that $A$ contains two contributions: (i) chemotactic alignment 
with the local gradient in $s$ and $p$ and (ii) slowing down of swimmers due to 
increased substrate consumption as a result of a local excess of $\rho$ that 
depletes $s$ locally. This provides 
a realization of the density-dependent self-propulsion velocity of \cite{fily}. 
In the overdamped limit, for large $A$ and $B$, one eigenmode with relaxation
}
\begin{eqnarray}\label{moseSc}
 && \hskip-1.2cm -i\omega =-D' q^2 \equiv   -(D + \frac{v_1^2 s_0^2}{6 D_r}
-\rho_0 \xi_s^2 B -\frac{v_1 s_0 \rho_0 \xi_s^2 A}{6 D_r}) q^2,
\end{eqnarray}
{goes unstable with growth rate $\sim q^2$ at small $q$.
Competition with stabilizing effects at larger $q$ will lead to a modulated
growth morphology with a length scale determined by the wavenumber of
peak growth $\sim |D'|^{1/2}$. The other mode, controlled by $D_r$, remains
stable for $q \to 0$, i.e., interactions do not promote flocking.
Working at large $D_r$ also justifies the overdamped limit.}

{In the parameter range where these modes are stable, the
steady-state static
small-$q$ structure factor takes the form $S_\rho \propto 1/{(D' + \gamma q^2)}$ where
\begin{eqnarray}\label{Srho}
&& \hskip-1.0cm \gamma = 2 \xi_s^2 D_r [\rho_0 \xi_s^2 B+\frac{\rho_0 \xi_s^2 A v_1 s_0}{6D_r}
+\frac{1}{3}\epsilon \rho_0 \xi_s^2 B (v_1 s_0- \rho_0 \xi_s^2 A)   \nonumber \\
&& \hskip-0.6cm + (D-\epsilon \rho_0 \xi_s^2 A) (D+\frac{2 v_1^2 s_0^2}{45 D_r}-\rho_0 \xi_s^2 B) \nonumber \\
&& \hskip-0.6cm + D' (\rho_0 \xi_s^2 B+\epsilon \rho_0 \xi_s^2 A -2D)].
\end{eqnarray}
For $\gamma>0$, as $D' \to 0^+$, $S_\rho$ displays fluctuations 
with a correlation length $\sqrt{\gamma/D'}$ that diverges as $D'\to 0$, presaging the onset
of clumping (see Fig. 3). Still in the linearly stable regime but
with $\gamma<0$, an
analysis to order $q^4$ shows that the system has
a tendency towards patterning with a characteristic length scale $\sim
|\gamma|^{-1/2}$, whose origin involves a competition between the
chemotactic ($A$) and phoretic ($B$) response to gradients.}

\section{Saturated}
Next we consider the saturated limit $\xi_s \to \infty$, realized by
working at saturation concentrations on the MM curve.
It is useful to define $\mathbf{E}(\mathbf{r}) = - \mN \int_{\mathbf{r}'}
\rho(\mathbf{r}')/|\mathbf{r}-\mathbf{r}'|$, which plays the role of an
electric field in Eq. \ref{saturated} below. Equation \ref{psscale}
then implies $\mN s= N \kappa_2 (\mathbf{E} - \epsilon \rho_0 \mP_L)/D_s$.
In this limit the orientation and density fields satisfy\footnote{In Eq. \ref{unsaturated}
we have displayed only those nonlinear terms required to stabilized a state of
nonzero $\mP_L$. The complete equations may be seen in the Supplementary Material.}
\begin{subequations}
\begin{eqnarray}\label{saturated}
 && \hskip-1.2cm [\partial_t - (D+\frac{v_0^2}{30 D_r}) \nabla^2] \mP_L+\frac{v_0}{3 \rho_0} \mN \rho + \frac{A}{3} \mathbf{E} \nonumber \\
 && \hskip-1.2cm - [\frac{v_0^2}{90 D_r} - \frac{2\epsilon  N \kappa_2 v_0 \rho_0}{135 D_r} (\frac{\beta_1}{D_p}+ \frac{\alpha_1}{D_s})] \mN \mN \cdot\mP_L     \nonumber \\
 && \hskip-1.2cm+[\frac{2 \epsilon^2 A^2 \rho_0^2 }{15 D_r} n_L^2  -  \frac{\epsilon A}{3} \rho_0 +2D_r] \mP_L= 0, \label{saturateda} \\
&& \hskip-1.2cm (\partial_t - D\nabla^2)\rho + \rho_0 (v_0 + \epsilon  \rho_0 B) \mN \cdot \mP_L  -
\rho_0 \Bb \mN \cdot \mathbf{E}=0, \label{saturatedb}
\end{eqnarray}
\end{subequations}
where $v_0 \equiv \lim_{s_0 \gg \kappa_2/\kappa_1} V_0(s_0)$.
The electric-field character of $\mathbf{E}$ is evident in
Eq. \ref{saturateda} through the alignment term $\propto A$
and the Ohmic current $\propto B$ in Eq. \ref{saturatedb}.
Note that a large and positive $\epsilon \rho_0 A/3$ can
destabilize the $n_{L} = 0$ state leading---once higher order terms are taken
into account---to a state of nonzero $\mP_L$, i.e. a condensation of asters. The
phenomenon is related to that reported in \cite{kripa2012}, with the important
difference in our case of long-range interactions mediated by $\mathbf{E}$, as
in \cite{thermophoretic2012}, with a resemblance to gravitational collapse
\cite{JeansInstability}.

In the overdamped limit, i.e. for sufficiently large $D_r$, the relaxation rates of the eigenmodes are
\begin{eqnarray}\label{saturatedmodes}
&&\hskip-1.2cm -i\omega =  \left \{ \begin{array}{c}
               \frac{G}{2D_r'} - [2 D+ \frac{v_0 (v_0 + \epsilon \rho_0 B)}{3 D_r'}] q^2,\\\\
            -2 D_r' {+ O(q^2)},
           \end{array} \right. 
\end{eqnarray}
where $D_r' = D_r - \epsilon \rho_0 A /6$ represents a modified rotational diffusion, and
$G = 2 {\rho_0 B D_r} + \frac{1}{3}\rho_0 A v_0$ is an effective control
parameter for the nature of interaction between the swimmers.
Equation \ref{saturated} shows that in the saturated limit the effective long-ranged
interaction between colloids (as mediated by S and P) leads to non-vanishing relaxation rates at $q=0$
for both modes, notwithstanding the conservation law governing $\rho$.

For $G<0$, the swimmers interact with long-ranged repulsive interactions.
The structure factor $S_\rho(q\to0) = 0 $, as it is a ratio of the strength
of fluctuations and the wavenumber independent relaxation rate, which is
reminiscent of suppression of charge density fluctuations in electrolytes.
Including terms of higher order in $q$ yields a density structure factor
with a peak at $q \sim G^{1/4}$, which is characteristic of
micro-phase separation (white region in Fig. 2). 
We see from Eq. \ref{saturated} that for $G>0$ and large $D_r'$ the isotropic
state with uniform density is linearly unstable for small wavenumber $q$, including $q=0$.
This effect
is a dissipative
analog of the gravitational Jeans instability \cite{JeansInstability}, and is a consequence
of the long-ranged attractive interaction (yellow). Related behavior has been predicted for thermophoretic
colloids \cite{thermophoretic2012}. Letting $D_r' \to 0$ by increasing $A$ and keeping
$G>0$ brings the system out of the overdamped region where the relaxation of
$\mn$ slows down and it behaves like a velocity field. This behavior 
where the system resembles a gravitational system conserving momentum and
displays an instability formally equivalent to the standard hydrodynamic Jeans
instability \cite{JeansInstability} (magenta). Modes with wavenumbers larger than a crossover
scale given by a competition between the interaction strength $G$ and a squared
sound speed equivalent $v_0 (v_0+\epsilon \rho_0 B)$ for our system are oscillatory,
whereas modes with smaller wavenumbers are too ``massive'' and collapse.
On further tuning the parameters to approach $D_r'<0$, notwithstanding the
value of $G$, one anticipates an instability towards a spontaneously oscillating
state (green). Restricting our attention to the stable case, we find a structure factor for
$\mP_L$ with a correlation length $\sim (D/D_r')^{1/2}$ that grows as $D_r'$
decreases, indicating strong fluctuations towards aster formation.
For small $D_r'$ and $G<0$ the response shows ``plasma oscillations'' \cite{plasma} with frequency
$\sim \sqrt{|G|}$. For $D_r'<0$ (dark green), 
the system can also develop spontaneous oscillations, or ringing.

\section{Summary}\label{summary}
A colloid patterned with catalyst and immersed in a maintained reactant 
medium is a minimal nonequilibrium particle, displaying directed motion 
and related behaviours ruled out at thermal equilibrium. We have 
determined theoretically {the nature of patterning that will cause such an
active colloid to reorient along} and move up or down a gradient of chemical
reactant, thus delineating the principles for the design of chemotactic 
self-phoretic particles. Coarse-graining the resulting Langevin equations 
for the position and polar axis of one particle, we discover the dynamics 
of the density and polar order parameter of a collection. The interplay of 
chemotaxis and phoresis leads to clumping and patterning at low reactant 
concentration; at high concentration, the slow decay of diffusing reactants 
and products yields analogues of electrostatic and gravitational 
phenomena -- Debye screening, microphase separation, plasma oscillations 
and gravitational collapse. The interactions promote aster formation, 
not a flocking transition, and the instabilities mediated by the 
long-range diffusion fields have a character distinct from those the 
generic instability driven by the velocity field in Stokesian active 
liquid crystals.  We look forward to experimental tests and, 
eventually, practical application of our predictions. 

\begin{acknowledgments}
RG and SR acknowledge HFSP grant
RGP0061/2013 and a J C Bose Fellowship respectively, and both thank
the Isaac Newton Institute for Mathematical Sciences where a part of this work
was completed. SS thanks TCIS for hospitality.
\end{acknowledgments}

\section{Appendix I: Expression for the sources}
\begin{figure}\label{Q_schem}
\begin{center}
\includegraphics[angle=0,width=3in]{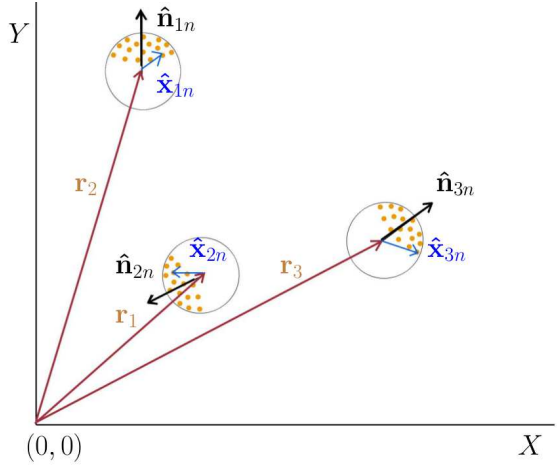}
\caption{Schematic for the calculation of $Q$.}
\label{phasedia1}
\end{center}
\end{figure}

We begin by obtaining an expression for the sources $Q\left(\mathbf{r},t\right)$ 
for generation of P and consumption of S in terms of the  coarse-grained 
density and orientation fields, 
$\sum_\alpha \delta\left(\mathbf{r} - \mathbf{r}_\alpha\right) = \rho\left(\mathbf{r}\right)$,
$\sum_\alpha \mathbf{\hat{n}}_\alpha \delta\left(\mathbf{r} - \mathbf{r}_\alpha\right) = \mathbf{w}\left(\mathbf{r}\right)$, 
of the colloids. $Q$ simply sums over all the enzymatic sites on the surfaces of all catalytic colloids. 
Setting $\mathbf{X}_\alpha = R \mathbf{\hat{x}}_\alpha$ (see Fig. \ref{Q_schem}), 
$\Omega_b$ as the infinitesimal solid angle and Taylor expanding upto $O\left(R^3\right)$ we get 
\begin{eqnarray}\label{sources}
Q\left(\mathbf{r},t\right) &=& \kappa(s)
\sum_{\alpha} \int_{|\mathbf{X}_{\alpha}| = R}
\delta(\mathbf{r}-\mathbf{r}_\alpha-\mathbf{X}_{\alpha}) \sigma
(\mathbf{X}_{\alpha} \cdot \mathbf{\hat{n}}_\alpha), \nonumber \\
&=& 
\kappa(s) \sum_{\alpha} \int_b R^2 \mbox{d} \Omega_b \left[\delta \left(\mathbf{r}-\mathbf{r}_\alpha\right)-R \mathbf{\hat{x}}_\alpha \cdot \mathbf{\nabla}\delta \left(\mathbf{r}-\mathbf{r}_\alpha\right) + \frac{R^2}{2} \left(\mathbf{\hat{x}}_\alpha \cdot \mathbf{\nabla}\right)^2 \delta \left(\mathbf{r}-\mathbf{r}_\alpha\right)\right] \nonumber \\
&& \times \left[\sigma_0 +\sigma_1 \mathbf{\hat{x}}_\alpha \cdot \mathbf{\hat{n}}_\alpha+ \sigma_2 \left(\mathbf{\hat{x}}_\alpha \cdot \mathbf{\hat{n}}_\alpha\right)^2\right]  \nonumber \\
&=& \kappa(s) 4 \pi R^2 \left[ \sigma_0 \rho - \frac{1}{3} \sigma_2 \rho- \frac{R \sigma_1 }{3} \mathbf{\nabla} \cdot \mathbf{n}\right] \nonumber \\
&=& \kappa(s) N \left[ \rho - \frac{\sigma_2}{3 \sigma_0}  \rho- \frac{R \sigma_1 }{3 \sigma_0} \mathbf{\nabla} \cdot \mathbf{n}\right],
\label{monodi}
\end{eqnarray}
where $N = 4 \pi R^2 \sigma_0$. The sources are obtained using 
\begin{eqnarray}
\int  \hat{x}_{i} \hat{x}_{j}   \mbox{d}\Omega_b  = \frac{4\pi}{3} \delta_{ij} 
\end{eqnarray}
\section{Appendix II: Derivation of the full nonlinear equations}
Next, starting from the Langevin equations, we construct
equations of motion for $\rho\left(\mathbf{r}\right)$ and $\mathbf{w}\left(\mathbf{r}\right)$. 
We construct the joint probability density $P$ for the position and orientation;
\begin{eqnarray}
 {P}\left(\mathbf{r},\mathbf{w}\right) = \langle \sum_{\alpha} \delta\left(\mathbf{r} - \mathbf{r}_\alpha\right) 
 \delta\left(\mathbf{w} - \mathbf{\hat{w}}_\alpha\right)\rangle
\end{eqnarray}
$P$ is used to derive the Fokker Planck equation for the joint distribution of particle position and polarity;
\begin{eqnarray}\label{FP}
 &&\partial_t P + \mathbf{\nabla} \cdot \left[P \left(V_0\left(s\right) \mathbf{\hat{w}} +
\alpha_0 \mathbf{\nabla}s + \alpha_1 \mathbf{\hat{w}} \mathbf{\hat{w}} \cdot \mathbf{\nabla}s + 
\beta_0 \nabla p + \beta_1 \mathbf{\hat{w}} \mathbf{\hat{w}} \cdot \mathbf{\nabla} p\right)\right] - D \nabla^2 P \nonumber \\
&& + \left(\mathbf{\hat{w}} \times \mathbf{\nabla}_{w}\right) \cdot \left[P\left(\Phi_0 \left(\mathbf{\hat{w}} \times
\mathbf{\nabla}s\right)\times \mathbf{\hat{w}} + \Omega_0 \left(\mathbf{\hat{w}} \times \mathbf{\nabla}
p\right)\times \mathbf{\hat{w}}\right)\right] - D_r \left(\mathbf{\hat{w}} \times \mathbf{\nabla_{w}}\right)^2 P = 0 
\end{eqnarray}
The equation for $\rho = \int P\left(\mathbf{r},\mathbf{\hat{w}}\right) d^2 w $ is  given by taking the zeroth moment of the above equations
\begin{eqnarray}
\partial_t \rho + \mathbf{\nabla} \cdot \left( V_0 \mathbf{w} \right) 
+ \mathbf{\nabla} \cdot \left[ \left(\beta_0 + \frac{\beta_1}{3}\right) \mathbf{\nabla} p
-\left(\alpha_0 + \frac{\alpha_1}{3} \right) \mathbf{\nabla} s  \rho 
+ \left(\beta_1 \mathbf{\nabla} p-  \alpha_1 \mathbf{\nabla}s \right) \cdot \bsf{Q}\right] -D \nabla^2 \rho = 0,
\label{rhoeq}
\end{eqnarray}
 where $D$ is the diffusivity of the swimmers and $\bsf{Q} = \int P\left(\mathbf{r},\mathbf{\hat{w}}\right) \left(\mathbf{\hat{w}}  \mathbf{\hat{w}} - \bsf{I}/3 \right) d^2 w$ is the nematic order parameter. 
The equation for the polar order parameter 
$\mathbf{w}\left(\mathbf{r}\right) = \int P\left(\mathbf{r},\mathbf{\hat{w}}\right) \mathbf{\hat{w}} d^2 w  $ is the first moment of \eqref{FP}
 \begin{eqnarray}
\partial_t \mathbf{w} &+& \mathbf{\nabla} \cdot \left(V_0  \bsf{Q}\right) + \frac{1}{3} \mathbf{\nabla} \left(V_0  \rho\right) + \mathbf{\nabla} \cdot \left[\left(-\alpha_0 \mathbf{\nabla} s 
+ \beta_0 \mathbf{\nabla} p\right)\mathbf{w}+ \left(-\alpha_1 \mathbf{\nabla} s + \beta_1 \mathbf{\nabla} p\right) \cdot \bsf{Q}^3\right]
  - D\nabla^2 \mathbf{w} \nonumber \\
&+ &\Phi_0 \left[\bsf{Q} \cdot \mathbf{\nabla}s - \frac{2}{3} \rho \mathbf{\nabla}s \right] + \Omega_0 \left[\bsf{Q} \cdot \mathbf{\nabla} p - \frac{2}{3} \rho \mathbf{\nabla} p \right]
+ 2D_r  \mathbf{w} = 0,
\label{eqN}
 \end{eqnarray}
where $\bsf{Q}^{3}  = \int P\left(\mathbf{r}, \mathbf{w}\right) \mathbf{\hat{w}}^3 d^2 w$. 
While coarse-graining, we see inevitably that the dynamical
equation for $\mathbf{w}$ involves higher moments of the orientational
distribution function which must be re-expressed in terms of lower moments
\cite{bertin2006,shelley,SMishra2013}. No process involving apolar
bundling of particles
is considered here, so we do not consider strict nematic order. The
occurrence of more than one factor of $\mathbf{\hat{w}}_\alpha$ can then 
lead only to higher powers of $\mathbf{w}$, combined with isotropic terms. 
We obtain a hierarchy of equation which have to be truncated using a closure condition. 

\subsection{Closure}
The condition often used is to assume that $\bsf{Q}$ relaxes fast so that 
$\partial_t \bsf{Q} = 0$, i.e. $\bsf{Q}$ is slaved to $\rho$ and $\bm{w}$. 
In this scheme of closure we systematically ignore terms higher than 
$|\mn|^3$ so that it holds close to the order disorder transition. Deep in 
the ordered state the closure scheme has to be modified keeping in mind that 
$\langle \mn \mn \rangle = w_0^2 \mathbf{\hat{k}} \mathbf{\hat{k}}$, where 
$\mathbf{\hat{k}}$ is the ordering direction and $w_0$ is magnitude, so that one has to look at fluctuations 
about this reference state.

We will now write down the equation of motion for the nematic order parameter and 
obtain an expression for $\bsf{Q}$ in terms of $\rho$ and $\mn$. 
The expression for $\bsf{Q}$ is substituted in the equations 
for $\rho$ and $\mn$ to obtain coarse grained 
equations of these variables in terms of $\rho$ and $\mn$ and involving 
the diffusion field $s$. Next we work in two different approximations 
to obtain expressions for $s$ in terms of $\rho$ and $\mn$ 
to obtain effective equations for $\rho$ 
and $\mn$. The nematic order parameter $\bsf{Q}$ satisfies the equation
\begin{eqnarray}
&& \partial_t \bsf{Q} + \mathbf{\nabla} \cdot \left[V_0\left( \bsf{Q}^{3} - \frac{\bsf{I}}{3} \mathbf{w}\right)\right] - \alpha_0 \mathbf{\nabla} \cdot \left( \mathbf{\nabla}s \bsf{Q} \right) - 
\alpha_1 \mathbf{\nabla} \cdot \left(\bsf{Q}^4 \cdot \mathbf{\nabla} s - \frac{\bsf{I}}{3} \bsf{Q}^{\prime} \cdot \mathbf{\nabla} s\right)  + \beta_0 \mathbf{\nabla} \cdot \left(\mathbf{\nabla}p \bsf{Q}\right)  \nonumber \\
&&
 +
\beta_1 \mathbf{\nabla} \cdot  \left(\bsf{Q}^4 \cdot \mathbf{\nabla} p - \frac{\bsf{I}}{3} \bsf{Q}^\prime \cdot \mathbf{\nabla} p\right) - D \nabla^2 \bsf{Q} + 
 2\Phi_0 \left(\bsf{Q}^{3} \cdot \mathbf{\nabla} s - \mathbf{w}\mathbf{\nabla}s\right) + 2\Omega_0 \left(\bsf{Q}^{3} \cdot \mathbf{\nabla}p - \mathbf{w}\mathbf{\nabla}p\right) 
\nonumber  \\ 
&&
+ 6 D_r \bsf{Q}= 0,
\label{eqQ}
\end{eqnarray}
where $\bsf{Q}^4  = \int P\left(\mathbf{r}, \mathbf{w}\right) \mathbf{\hat{w}}^4 d^2 w$ and $\bsf{Q}'  = \int P\left(\mathbf{r}, \mathbf{w}\right) \mathbf{\hat{w}}^2 d^2 w$.\\
The traceless parts of higher moments like $\bsf{Q}^3$ and $\bsf{Q}^4$ are assumed to be negligibly small, so that
\begin{eqnarray}\label{moments}
 {Q}^{3}_{\alpha \beta \gamma} &\to& \frac{1}{5}\left(w_\alpha \delta_{\beta \gamma} + w_\beta \delta_{\gamma \alpha} + w_\gamma \delta_{\alpha \beta}\right) \nonumber \\
 {Q}^{4}_{\alpha \beta \gamma \mu} &\to& \frac{1}{15}\left(\delta_{\alpha \beta} \delta_{\gamma \mu} + 
\delta_{\alpha \gamma} \delta_{\beta \mu} + \delta_{\alpha \mu} \delta_{\gamma \beta} \right)
\end{eqnarray}
Using \eqref{moments} and $p = - {s D_s}/{D_p}$, and defining for any two vectors $\mathbf{v}_1$ and $\mathbf{v}_2$, 
\begin{eqnarray}
\left(\mathbf{v}_1 \mathbf{v}_2\right)^{TS} = \frac{1}{2}\left(\mathbf{v}_1 \mathbf{v}_2+\mathbf{v}_2 \mathbf{v}_1\right) - \left(\mathbf{v}_1 \cdot \mathbf{v}_2\right)\frac{\bsf{I}}{3},
\end{eqnarray} 
and keeping terms upto the lowest order in gradients we have
 \begin{eqnarray}
 \bsf{Q} &=& \frac{1}{15 D_r}\left[-\mn \mN V_0- V_0 \mN \mn + \frac{C_3}{3} \rho \mN\mN s + \frac{C_3}{3} \mN \rho \mN s - 3 C_2 \mn \mN s\right]^{TS},
 \label{Qsol}
 \end{eqnarray}
where 
\begin{eqnarray}
C_1 &=& \alpha_0 + \frac{\alpha_1}{5} + \frac{D_s}{D_p}\left(\beta_0 + \frac{\beta_1}{5}\right) \nonumber \\
C_2 &=& \frac{D_s}{D_p} \Omega_0 - \Phi_0 \nonumber \\
C_3 &=& \alpha_1 + \frac{D_s}{D_p} \beta_1 .
\end{eqnarray} 
We also define 
\begin{eqnarray}
B &=&  N \kappa(s)\left[\frac{1}{D_s}\left( \alpha_0 + \frac{\alpha_1}{3}
\right) + \frac{1}{D_p} \left(\beta_0 +
\frac{\beta_1}{3}\right)\right],\nonumber \\
A &=& N \kappa(s) [\frac{\Phi_0}{D_s} - \frac{\Omega_0}{D_p} + \Delta
\frac{V_0(s_0)}{2 D_s}],
\end{eqnarray}
where $\Delta = d \ln\kappa(s)/ds|_{s_0}$. 
\subsection{Saturated: Equation of Motion for $\mn$}
In this regime, the phoretic velocity is independent of the substrate concentration so that $\mN V_0 = 0$ 
and we substitute $\lim_{s_0 >> \kappa_2/\kappa_1} V_0\left(s\right) \to v_0$ and $\lim_{s_0 >> \kappa_2/\kappa_1} \kappa(s_0) \to \kappa_2$. 
Substituting $\bsf{Q}$ from \eqref{Qsol} into \eqref{eqN} and 
using
\begin{eqnarray}\label{Qform}
 \bsf{Q}^3\cdot \mN s = \frac{1}{5}\left(\mn \mN s+ \mN s \mn + \left(\mn \cdot \mN s\right) \bsf{I}\right)
\end{eqnarray}
we have the equation of motion for the polar order parameter
\begin{eqnarray}\label{NS}
&&  \partial_t \mn  - \left(D+ \frac{v_0^2}{30 D_r}\right) \nabla^2 \mn - \frac{v_0^2}{90D_r} \mN\left(\mN \cdot \mn\right) + 2D_r \mn +\frac{2 C_2}{3} \rho \mN s + \frac{v_0}{3} \mN \rho   \nonumber \\
&&  - \left(\frac{11 C_2 v_0}{90D_r} + \frac{C_3}{5}\right) \left(\mN \cdot \mn\right) \mN s  - \left(\frac{C_2 v_0}{10D_r} + \frac{C_3}{5}\right) \left(\mn \cdot \mN\right)\mN s - \left(\frac{C_2 v_0}{10D_r} +C_1\right)\left(\nabla^2 s\right) \mn  \nonumber \\
&&  - \left(\frac{C_2 v_0}{15 D_r} +C_1\right)\left(\mN s \cdot \mN\right)\mn- \left(\frac{C_3}{5} - \frac{C_2 v_0}{15D_r}\right) \mN\left(\mn \cdot \mN s\right) + \frac{C_2 v_0}{30D_r}\left(\mN \mn\right)\cdot \mN s \nonumber \\
&&  + \frac{C_3 v_0}{270 D_r} \left[ 4 \rho \mN\left(\nabla^2s\right) + 3 \left(\mN s \cdot \mN\right) \mN \rho - 2 \mN\left(\mN \rho \cdot \mN s\right) + 9 \left(\mN \rho \cdot \mN\right) \mN s + 3 \nabla^2 \rho \mN s + \left(\nabla^2 s\right) \mN \rho \right] \nonumber \\
&&  +\frac{C_2^2}{10 D_r} \mn |\mN s|^2 + \frac{C_2^2}{30D_r}\left(\mn \cdot \mN s\right)\mN s \nonumber \\
&&  + \frac{C_2C_3}{270D_r} \left[ 2 \rho \left(\nabla^2s\right)\mN s - 6 \rho\left(\mN \mN s\right)\cdot \mN s - 3 \mN \rho |\mN s|^2  -  \mN s \left(\mN s \cdot \mN \rho\right) \right]=0
\end{eqnarray}
In the saturated limit we have $ \mathbf{\nabla}s = N \kappa_2 \left(\bm{E} - \epsilon \mathbf{w}_L \right)/D_s$ and 
$\nabla^2 s = N \kappa_2 \left(\rho - \epsilon \mathbf{\nabla} \cdot \mathbf{w}\right)/D_s$. We have defined the curl free or 
longitudinal component of 
$\mn$ expressed in the fourier space as $\mn_{\mathbf{q} L} = \mathbf{\hat{q}} \mathbf{\hat{q}} \cdot \mn_{\mathbf{q}}$. 
In what follows we use $p = -{s D_s}/{D_p}$ and define 
$c_{1,3} = N \kappa_2 C_{1,3}/D_s$. 
The effective
equation for $\mathbf{w}$ is then 
\begin{eqnarray}\label{NE}
&& \hskip-1cm \partial_t \mn - \left(D+ \frac{v_0^2}{30 D_r}\right) \nabla^2 \mn + \left(\frac{2\epsilon  c_3 v_0}{135 D_r} \rho 
- \frac{v_0^2}{90 D_r}\right) \mN \left(\mN \cdot \mn\right) + \left(\frac{v_0}{3} + f_1 \right) \mN \rho \nonumber \\
&& \hskip-1cm + \left( 2 D_r + f_2 \right) \mn  + \left(\frac{A^2 \epsilon^2 }{30 D_r} |\mn_L|^2 - \frac{2 A \epsilon}{3} \rho 
+ f_3 \right) \mn_L + \left(\frac{2 A}{3} \rho  - f_4  \right) \mathbf{E}  \nonumber \\
&& \hskip-1cm + \mN \Pi + [\mn \mN \mathbf{E}] + [\mn \mn \mathbf{E}] + [(\mathbf{E}-\epsilon \mn_L) \mN \rho]  + [\mn \mN \mn]=0,
\end{eqnarray}
where the square brackets denote a linear combination of all possible contractions 
of the enclosed terms to obtain a vector. The equation is asymptotically exact as $\xi_s \to \infty$. 
We will now discuss the significance of various 
terms and provide expressions for the functions 
introduced above. As the swimmer 
moves with an average speed $v_0$ along its symmetry axis, the polar order parameter 
is equivalent to a velocity field. With this interpretation $\rho$ becomes a pressure 
like term where corrections to the squared sound speed $v_0/3$ is given by 
\begin{eqnarray}
 f_1 = -\frac{v_0 c_3}{270D_r} \delta \rho  - \frac{\epsilon v_0 c_3}{90 D_r}  \left( \mN \cdot \mn_L\right). \nonumber
\end{eqnarray}
Similarly $\Pi$ is a generalised pressure term depending on the other fields as
\begin{eqnarray}
\Pi =  (\mathbf{E} - \epsilon \mn_L) \cdot\left[\left(\frac{v_0 A}{15 D_r} - \frac{c_3}{5} +  \frac{\epsilon A c_3}{90 D_r} \rho \right) \mn_L
 - \frac{v_0 c_3}{135 D_r} \mN \rho - \frac{A c_3}{90 D_r} \rho \mathbf{E} \right]. \nonumber
\end{eqnarray}
The presence of the swimmer creates spatial variation in the S and P. This is encoded in the fields 
$\mathbf{E}$ and $\mn_L$ of which the first enters \eqref{NE} as an external orienting field while the 
second enters the dynamics as a spontaneous symmetry breaking effect where the linear term $(D_r - 2 \rho A/3)$ can 
change sign producing $\mn_L$ condensation. There are higher order stabilising terms with coefficient $\propto A^2$ which restrain the 
growth of $|\mn_L|$. The linear terms coefficients are modified by the corrections
\begin{eqnarray}
 && f_2 = \frac{\epsilon^2 A^2 }{10 D_r} |\mn_L|^2 + \delta \rho\left(c_1 + \frac{A v_0}{10 D_r}\right) + \frac{A^2}{10 D_r} |\mathbf{E}|^2, \nonumber \\
 && f_3 = \frac{\epsilon A c_3}{45 D_r} \left[  \frac{1}{6}\mN \rho \cdot (\mathbf{E} - \epsilon \mn_L) - \rho \delta \rho  \right], \nonumber \\
 && f_4 = \frac{A c_3}{135  D_r} \rho \delta \rho + \frac{ \epsilon^2 A^2 }{30 D_r} |\mn_L|^2 + \frac{A c_3}{270 D_r} \mN \rho \cdot\left(\mathbf{E} - \epsilon \mn_L\right),
\end{eqnarray}
which can destabilise the condensed state. Note that the terms that 
dictate the nature of the aster condensation are all proportional 
to the polarity of the catalytic coat as they are sensitive to the 
polar nature of the diffusion field.

The advective nonlinearities have the same structure as in the Toner-Tu model \cite{TonerTu1995} 
given that $\mn$ is like a velocity that can carry other fields. However note the absence of terms of the form $(\mn \cdot \mN)\mn$; this 
is because the interaction in the two cases are very different and in this particular closure that holds close 
to the transition the average value of $\mn$ and thus the coefficients of advective terms are proportional to the 
interaction strength which in this case is $\epsilon$. Bertin et. al. \cite{bertin2006} modelled the Toner-Tu 
like interaction as an actual binary collision which tends to align the particles. In this system where interactions are mediated 
by the diffusion fields alone, the interaction is mediated by the $\mn_L$ term as discussed above. 
\begin{eqnarray}\label{SatAdvc}
\hskip-1cm \left[\mn \mN \mn \right] &\equiv& \epsilon  \left(\frac{ c_3}{5} - \frac{A c_3  \epsilon}{135 D_r} \rho  + \frac{11 A v_0 }{90 D_r}\right) \left(\mN \cdot \mn\right) \mn_L + \epsilon\left(\frac{A v_0 }{10 D_r} + \frac{c_3}{5} \right) \left(\mn \cdot \mN\right) \mn_L   \nonumber \\
&& \hskip-1cm - \epsilon \left(\frac{A v_0}{10 D_r} + c_1\right) \mn \left(\mN \cdot \mn\right) + \epsilon \left(\frac{A v_0}{15 D_r} + c_1\right) \left(\mn_L \cdot \mN\right) \mn - \epsilon \frac{A v_0 }{30 D_r} \mn_L \times \left(\mN \times \mn\right).  \nonumber
\end{eqnarray}
The other nonlinearities are the terms coupling $\mn$ and $\mathbf{E}$ which to lowest order are
\begin{eqnarray}
\left[ \mn \mN \mathbf{E} \right] &\equiv& \left(\frac{ c_3}{5} - \frac{A c_3  \epsilon}{135 D_r} \rho  + \frac{11 A v_0 }{90 D_r}  \right) \left(\mN \cdot \mn\right)\mathbf{E} + \left(c_1  +  \frac{A v_0 }{15 D_r}\right)\left(\mathbf{E} \cdot \mN\right) \mn    \nonumber \\
&& + \left(\frac{ c_3}{5} + \frac{A v_0 }{10 D_r}\right)\left(\mn \cdot \mN\right)\mathbf{E}+ \frac{A v_0}{30 D_r} \mathbf{E} \times \left(\mN \times \mn\right) , \nonumber
\end{eqnarray}
and higher order terms
\begin{eqnarray}
\hskip-1cm \left[\mn \mn \mathbf{E} \right] &\equiv& - \frac{\epsilon A^2 }{45 D_r} \mn \left(\mn_L \cdot \mathbf{E}\right) - \frac{\epsilon A^2 }{30 D_r} \mn_L \left(\mn \cdot \mathbf{E}\right) + \frac{A^2}{30 D_r} \left(\mn \cdot \mathbf{E}\right)\mathbf{E}. \nonumber
\end{eqnarray}
The rest of the nonlinearities couple $\mN \rho$ and $(\mathbf{E}-\epsilon \mn_L)$:
\begin{eqnarray}
&& \hskip-1cm [\mN \rho (\mathbf{E}-\epsilon \mn_L)]= \frac{v_0 c_3}{90 D_r} \left[(\mathbf{E}-\epsilon \mn_L) \cdot \mN \right]\mN \rho
+ \frac{v_0 c_3}{30 D_r} \left(\mN \rho \cdot \mN\right)\left(\mathbf{E} - \epsilon \mn_L\right) + \frac{v_0 c_3}{90 D_r} \left(\mathbf{E} - \epsilon \mn_L\right)\nabla^2 \rho. \nonumber
\end{eqnarray}
We linearise \eqref{NE} to obtain
\begin{eqnarray}\label{Nlin}
&&\partial_t \mn - \left(D+ \frac{v_0^2}{30 D_r}\right) \nabla^2 \mn + \left(\frac{2\epsilon  c_3 v_0 \rho_0}{135 D_r} - \frac{v_0^2}{90 D_r}\right) \mN \left(\mN \cdot \mn\right) + \left(\frac{v_0}{3} - \frac{2 c_3 v_0 \rho_0}{135 D_r}\right) \mN \rho  + 2 D_r  \mn  - \frac{2 A \epsilon}{3} \rho_0  \mn_L \nonumber \\
&&  + \frac{2 A}{3} \rho_0   \mathbf{E}=0
\end{eqnarray}
Defining $\mm$ by $\mn/\rho_0$; and the transverse component of $\mm$ as 
$\mm_{\mathbf{q}t} = \left(\bsf{I} - \hat{\mathbf{q}} \hat{\mathbf{q}}\right) \cdot \mm_{\mathbf{q}}$ we get
\begin{eqnarray}
&&\left[\partial_t - \left(D+ \frac{v_0^2}{30 D_r}\right) \nabla^2 - \left( \frac{v_0^2}{90 D_r} - \frac{2\epsilon  c_3 v_0 \rho_0}{135 D_r}\right) \mN \mN \cdot\right] \mm_L + \left( \frac{4\epsilon^2 A^2 }{30 D_r} + 2 D_r- \frac{2 A \epsilon \rho_0}{3}\right) \mm_L\nonumber \\
&&+ \rho_0^{-1}\left(\frac{v_0}{3} - \frac{2 c_3 v_0 \rho_0}{135 D_r}\right) \mN \rho  + \frac{2 A}{3}   \mathbf{E}=0 \label{linNl}\\
&&\left[\partial_t - \left(D+ \frac{v_0^2}{30 D_r}\right) \nabla^2 + \left(\frac{2\epsilon  c_3 v_0 \rho_0}{135 D_r} - \frac{v_0^2}{90 D_r}\right) \mN \mN \cdot\right] \mm_t + \left( \frac{\epsilon^2 A^2 \rho_0^2}{10 D_r} \mm_L^2 + 2 D_r\right) \mm_t = 0
\end{eqnarray}
Note that for $D_r>0$, $\mm_t$ stays in the overdamped region and is a fast degree of freedom that we throw away. 
Fourier transforming Eq.\eqref{linNl} we get 
\begin{eqnarray}
 \left[\partial_t + (D + \frac{v_0^2}{90D_r} - \frac{2 \epsilon N \kappa_2 v_0 \rho_0}{135 D_r}(\frac{\alpha_1}{D_s}+\frac{\beta_1}{D_p})) q^2 + 2(D_r - \frac{\epsilon \rho_0 A}{3})\right] i\mathbf{q} \cdot \mm_{l\mathbf{q}} - \frac{v_0}{3 \rho_0} q^2 \rho_{\mathbf{q}} + \frac{2 A}{3} \rho_{l\mathbf{q}} = 0,
\end{eqnarray}
\subsection{Saturated: Equation of Motion for $\rho$}
Substituting the closed form for $\bsf{Q}$ from Eq. \eqref{Qform} into Eq. \eqref{rhoeq} we 
get the equation for $\rho$ 
\begin{eqnarray}\label{rhoS}
&&\partial_t \rho + v_0 \mN \cdot \mn - \frac{B D_s}{\kappa_2 N} \mN\cdot \left(\rho \mN s\right)  -D \nabla^2 \rho \nonumber \\
&& - \frac{C_3}{15D_r} \mN \cdot \left[-\frac{v_0}{2}\left(\mN \mn\right) \cdot \mN s -\frac{v_0}{2}\left(\mN s \cdot \mN\right)\mn + \frac{v_0}{3} \left(\mN \cdot \mn\right) \mN s - \frac{A}{2} \left(\mN s\cdot \mn\right) \mN s \right. \nonumber \\
&&\left. + \frac{C_3}{3 } \rho \left(\mN \mN s\right)\cdot \mN s - \frac{C_3}{9} \rho \left(\mN s\right)\nabla^2 s + \frac{C_3}{3} \mN \rho |\mN s|^2 + \frac{C_3}{6} \mN \rho \left(\mN s \cdot \mN \rho\right) \right] =0
\end{eqnarray}
Substituting for $\mN s$ and using $c_{1,3} = N \kappa_2 C_{1,3}D_s^{-1}$ we have
\begin{eqnarray}\label{rhoE}
&& \partial_t \rho + v_0 \mN \cdot \mn -D \nabla^2 \rho  
-  \mN\cdot \left[ \mathbf{E} \left(B \rho + \frac{\epsilon c_3 A}{30 D_r} w_L^2 + \frac{c_3^2}{135 D_r} \rho \delta \rho \right) \right] \nonumber \\
&& + \epsilon  \mN \cdot \mn_L \left(B \rho + \frac{\epsilon c_3^2}{135 D_r} \rho \delta \rho + \frac{\epsilon^2 c_3 A}{30 D_r} w_L^2 \right) -\frac{c_3}{15D_r}\mN \cdot \mathcal{J}  = 0,
\end{eqnarray}
where $\delta \rho = \rho - \rho_0$ and $\mathcal{J}$ is the nonlinear part of the current:
\begin{eqnarray}
&& \hskip-1cm \mathcal{J} = \left(\frac{\epsilon c_3}{9} \rho - \frac{v_0}{3}\right)\left(\mN \cdot \mn\right)\mathbf{E} + v_0 \left(\mathbf{E}\cdot\mN\right)\mn  + \frac{v_0}{2} \mathbf{E} \times \left(\mN \times \mn\right)  + \frac{A}{2} \left(\mn \cdot \mathbf{E}\right)\mathbf{E} - \frac{A \epsilon}{2}\left(\mn \cdot \mathbf{E}\right)\mn_L \nonumber \\
&& \hskip-1cm - \epsilon v_0 \left(\mn_L \cdot \mN\right)\mn+ \left(\frac{\epsilon v_0}{3} - \frac{\epsilon^2 c_3}{9} \rho\right)\left(\mN\cdot \mn\right)\mn_L  - \frac{v_0 \epsilon}{2} \mn_L \times \left(\mN \times \mn\right) \nonumber \\
&& \hskip-1cm + \frac{c_3}{3} \mN \rho |\mathbf{E} - \epsilon \mn_L|^2 + \frac{c_3 D_s}{6 N \kappa_2} \mN \rho \left(\mN \rho \cdot \mathbf{E} - \epsilon \mN \rho \cdot \mn_L\right)
\end{eqnarray}
Eq \eqref{rhoE} is linearised and written in terms of $\mm$:
\begin{eqnarray}\label{rholina}
 \left(\partial_t -D \nabla^2\right) \rho + \rho_0\left(v_0 + \epsilon B \rho_0\right) \mN \cdot \mm_L - B \rho_0 \mN\cdot \mathbf{E}  = 0
\end{eqnarray}
Fourier transforming we get
\begin{eqnarray}
 \left(\partial_t + D q^2 - B \rho_0\right) \rho_{\mathbf{q}} + \rho_0\left(v_0 + \epsilon B \rho_0\right) i \mathbf{q} \cdot \mm_{L \mathbf{q}} = 0
\end{eqnarray}

\subsection{Unsaturated: equation of motion for $\mn$}
In the linear part of the MM curve where $\kappa\left(s\right) = \kappa_1 s$, $V_0$ varies linearly with substrate concentration so that 
we can define the self phoretic velocity as $V_0\left(s\right) = \vl s$. The equation for $\mn$ in the unsaturated limit is 
\begin{eqnarray}
&& \partial_t \mn - \left(D+\frac{\vl^2 s^2}{30 D_r}\right)\nabla^2 \mn - \frac{\vl^2 s^2}{90D_r}\mN\left(\mN \cdot \mn\right) + \frac{2 C_2}{3} \rho \mN s + \frac{\vl}{3}  \mN(s \rho) + 2D_r \mn \nonumber \\
&& + \left[\frac{\left(\vl - 11C_2\right)\vl}{90D_r}s-\frac{C_3}{5}\right]\left(\mN \cdot \mn\right) \mN s  - \left[\frac{\left(C_2 + 5 \vl\right)\vl}{30D_r}s + C_1\right]\left(\mN s \cdot \mN\right)\mn - \left[\frac{\left(3C_2 + \vl\right)\vl}{30D_r}s + \frac{C_3}{5}\right]\left(\mn \cdot \mN\right) \mN s  \nonumber \\
&& - \left[\frac{\left(3C_2+\vl\right)\vl}{30D_r} s+C_1\right] \mn \nabla^2 s + \left[\frac{\left(3C_2 + \vl\right)\vl}{45 D_r}s-\frac{C_3}{5}\right]\mN\left(\mn \cdot \mN s\right) -\left[\frac{\left(2 \vl - C_2\right)\vl}{30 D_r}\right]\mN s \times \left(\mN \times \mn\right) \nonumber \\
&& + \frac{2C_3 \vl}{135D_r} s \nabla^2\left(\mN s\right) - \frac{\left(\vl - C_2\right)\left(3C_2 + \vl\right)}{90 D_r} \left[ \mn |\mN s|^2 + 3 \mN s \left(\mn \cdot \mN s\right) \right] \nonumber \\
&& + \frac{C_3\left(\vl-C_2\right)}{90D_r} \left[\rho \mN |\mN s|^2 - \frac{2}{3} \rho \nabla^2 s \mN s \right]  + \frac{\left(C_2+\vl\right) C_3}{90 D_r}\left[ \frac{1}{3} \mN s \left(\mN s \cdot \mN \rho\right) +  \mN \rho |\mN s|^2  \right] \nonumber \\
&& + \frac{C_3 \vl s}{270 D_r} \left[ 9 \left(\mN \rho \cdot \mN\right) \mN s  - 2 \left(\nabla^2 s\right) \mN \rho +  \left(\nabla^2 \rho\right) \mN s + 3 \left(\mN s \cdot \mN\right)\mN \rho - 2 \mN\left(\mN \rho \cdot \mN s\right) \right]= 0
\end{eqnarray}
To obtain the effective equation for $\mn$ we substitute $s \to s_0 + \delta s$; in the highly unsaturated 
limit we have $\delta s = - \xi_s^2 N \kappa(s_0) \left(\delta \rho - \epsilon \mN \cdot \mn\right) / D_s$ for $q<<\xi_s^{-1}$. 
Keeping terms upto second order in gradients we get
\begin{eqnarray}
&&\partial_t \mn - \left(D+\frac{\vl^2 s_0^2}{30 D_r}-\frac{2 N \xi_s^2 \vl^2 s_0 \kappa}{30 D_r D_s} \delta \rho\right)\nabla^2 \mn + \left(\frac{ \xi_s^2 A }{3} \rho -\frac{\vl^2 s_0^2}{90D_r} 
+ \frac{2 N \kappa \xi_s^2 \vl^2 s_0}{90D_r} \delta \rho\right)\mN\left(\mN \cdot \mn\right) + 2D_r \mn \nonumber \\
&& + \mN \rho \left[ \frac{\vl s_0}{3} - \frac{\xi_s^2 A}{3} \rho  - \frac{N \kappa \xi_s^2 \vl}{3} \left(\delta \rho - \epsilon \mN \cdot \mn\right) \mN \rho \right]  \nonumber \\
&&  - \frac{\xi_s^4 (3 \vl N \kappa  - A D_s)(3A D_s - \vl N \kappa )}{360 D_r D_s^2} \left[ 3 \mn |\mN \rho|^2 + \mN \rho(\mn \cdot \mN \rho)\right]  \nonumber \\
&& + \xi_s^2 [\mN \rho \mN \mn]= 0.
\end{eqnarray}
The pressure is modified due to two separate contributions: (1) because of chemotactic alignment of swimmers with the  local 
chemical gradient through $\rho \mN s$ and (2) due to the change in self phoretic 
velocity with change in background substrate concentration through the term $\mN(s \rho)$. The advective nonlinearities are
\begin{eqnarray}\label{UnsatAdvc}
&& [\mN \rho \mN \mn] = \nonumber \\
&&- \left[\frac{\left(13 \vl N \kappa  - 11 A D_s \right)\vl s_0}{180D_r D_s}  - \frac{N \kappa C_3}{5 D_s}\right]\left(\mN \cdot \mn\right) \mN \rho + \left[\frac{\left(A D_s + 9\vl N \kappa \right)\vl s_0}{60D_r D_s}+ \frac{N \kappa C_1}{D_s} \right]\left(\mN \rho \cdot \mN\right)\mn \nonumber \\
&& + \left[\frac{\left(3A D_s -\vl N \kappa \right)\vl s_0}{60D_r D_s} + \frac{C_3 N \kappa}{5 D_s}\right]\left(\mn \cdot \mN\right) \mN \rho  + \left[\frac{\left(3A  D_s -\vl N \kappa \right)\vl s_0}{60D_r D_s}  + \frac{N \kappa C_1}{D_s} \right] \mn \nabla^2 \rho \nonumber \\
&& + \left[\frac{C_3 N \kappa}{5  D_s} - \frac{\left( 3A  D_s -\vl N \kappa \right)\vl s_0 }{90 D_r D_s} \right] \mN\left(\mn \cdot \mN \rho\right) + \frac{\left(5 \vl N \kappa  - A D_s\right)\vl }{60 D_r D_s}\mN \rho \times \left(\mN \times \mn\right).
\end{eqnarray}
Note that the advective nonlinearities given in Eqs. \eqref{SatAdvc} and \eqref{UnsatAdvc} 
are very different in the two regimes because in the unsaturated limit 
$\mn_L$ acts like an ordering field while in the other the local chemical gradient 
is given by $\mN(\mN \cdot \mn)$. The linearised equation for $\mn$ is 
\begin{eqnarray}
&& \hskip-1cm \partial_t \mn - \left(D+\frac{\vl^2 s_0^2}{30 D_r}\right)\nabla^2 \mn - \frac{\vl^2 s_0^2}{90D_r}\mN\left(\mN \cdot \mn\right) - A \xi_s^2 \rho_0 \mN \left(\rho - \epsilon \mN \cdot \mn\right) + \frac{\vl s_0}{3} \mN \rho  + 2D_r \mn = 0.
\end{eqnarray}
The equation for $\mm_L$ and $\mm_t$ are
\begin{eqnarray}
&& \left[\partial_t + 2D_r - \left(D+\frac{\vl^2 s_0^2}{30 D_r}\right)\nabla^2 - \frac{\vl^2 s_0^2}{90D_r}\mN \mN \cdot\right] \mm_L + \frac{\epsilon  \rho_0 \xi_s^2 A}{3} \mN \left(\mN \cdot \mm_L\right)  + \left[\frac{\vl s_0}{3 \rho_0} - \frac{ \xi_s^2 A }{3 }\right] \mN \rho = 0, \nonumber \\
&& \left[\partial_t + 2D_r - \left(D+\frac{\vl^2 s_0^2}{30 D_r}\right)\nabla^2 \right]\mm_t = 0
\end{eqnarray}
Fourier transforming the equation for $\mn_L$ get
\begin{eqnarray}
\left[\partial_t + 2D_r + (D+\frac{2\vl^2 s_0^2}{45 D_r}) q^2\right] i\mathbf{q} \cdot \mn_{L\mathbf{q}} - q^2\frac{\left[{\vl s_0} - \rho_0 A \xi_s^2 \right]}{3 \rho_0} \rho_{\mathbf{q}} - q^2 \frac{\epsilon \rho_0 A}{3} i\mathbf{q} \cdot \mn_{L\mathbf{q}}= 0, 
\end{eqnarray}

\subsection{Unsaturated: equation of motion for $\rho$}
The equation for $\rho$ is
\begin{eqnarray}
&& \partial_t \rho - D \nabla^2 \rho + \vl s \left(\mN \cdot \mn \right) - \frac{B D_s}{N \kappa} \rho \left(\nabla^2s \right) 
- \frac{B D_s}{N \kappa} \left(\mN s \cdot \mN\right)\rho + \frac{C_3}{45 D_r} \mN \cdot \mathcal{J}(\mN s,\mN \rho,\rho,s) = 0,
\end{eqnarray}
where $\mathcal{J}$ is the nonlinear part of the current. Substituting for $\delta s$, the effective equation for $\rho$ upto $O\left(\nabla^3\right)$ is
\begin{eqnarray}
 && \partial_t \rho - (D - B \xi_s^2 \rho) \nabla^2 \rho - \epsilon B \xi_s^2 \rho \nabla^2 (\mN \cdot \mn) 
 +  B \xi_s^2 \left[|\mN \rho|^2 - \epsilon \mN \rho \cdot \mN\left(\mN \cdot \mn\right)\right]    \nonumber \\
 && + \mN \cdot \mn \left[\vl s_0  - \frac{\vl N \kappa \xi_s^2}{D_s} \left(\delta \rho - \epsilon \mN \cdot \mn\right) \right] 
 + \frac{N \kappa C_3}{45 D_r D_s}\mN \cdot \mathcal{J} = 0,
\end{eqnarray}
with the current
\begin{eqnarray}
 &&\mathcal{J} =  \frac{\xi_s^2 \left(3A D_s - \vl N \kappa \right)}{4 D_s} \left[ \mn|\mN \rho|^2 
 + \frac{1}{3}\mN \rho \left(\mn \cdot \mN \rho\right) \right] - \vl \left(s_0 - \frac{N \xi_s^2  \kappa}{D_s} 
 \delta \rho\right)\left(\mN \rho \cdot \mN\right)\mn   \nonumber \\
 &&  - \frac{\vl }{2} \left(s_0 - \frac{N \xi_s^2 \kappa }{D_s} \delta \rho\right) \mN \rho \times\left(\mN \times \mn\right) 
 - \frac{\vl}{3} \left(s_0 - \frac{N \xi_s^2  \kappa}{D_s} \delta \rho\right) \left(\mN \cdot \mn\right)\mN \rho  = 0
\end{eqnarray}
The linearised equation for $\rho$ written in terms of $\mm$ is
\begin{eqnarray}
 && \left[\partial_t - \left(D - \xi_s^2 B \rho_0\right) \nabla^2\right] \rho_0^{-1} \rho + \left(\vl - \epsilon \xi_s^2 B \rho_0 \nabla^2\right) \mN \cdot \mm = 0
\end{eqnarray}

\section{Appendix III: Linear analysis and Mode structure in the saturated regime : details }
In the saturated limit the dynamical matrix takes the form
\begin{eqnarray}\label{dynaS}
\bsf{M}_s= \left[ \begin{array}{cc}
            Dq^2-\rho_0 B  & \rho_0 (v_0 + \epsilon \rho_0 B)  \\
               \rho_0 \Aa/3 - v_0q^2/3 \rho_0 & (D + \frac{v_0^2}{30 D_r}) q^2 + 2D_r'\\
            \end{array}
 \right],
\end{eqnarray}
where $D_r' \equiv D_r - \epsilon \rho_0 \Aa /6$. The mode structure is
\begin{eqnarray}\label{saturatedmodes}
&&\hskip-1.2cm -i\omega =  \left \{ \begin{array}{c}
               \frac{G}{2D_r'} - [2 D+ \frac{v_0 (v_0 + \epsilon \rho_0 B)}{3 D_r'}] q^2,\\\\
            -2 (D_r-\frac{\rho_0 B}{2}) - [2 D+\frac{2 v_0^2}{45 D_r} - \frac{2\epsilon  N \kappa_2 v_0 \rho_0}{135 D_r} (\frac{\beta_1}{D_p}+ \frac{\alpha_1}{D_s})] q^2,
           \end{array} \right. \nonumber \\
&& \hskip-1.2cm\mbox{where           } G = {2\rho_0 B D_r} + \frac{1}{3}\rho_0 A v_0.
\end{eqnarray}
Note that $G$ gives the strength and sign of effective long ranged 
interaction between the centres of mass of two swimmers taking into account both phoretic and chemotactic 
response of swimmers to S and P gradients. We see in \eqref{saturatedmodes} that 
the main novelty of the saturated limit is that it leads to a non-vanishing relaxation 
rate at $q=0$ for both modes, notwithstanding the conservation law governing $\rho$. 
For $D_r>>0$, there is a wide separation in time scale between the two modes; the first mode whose corresponding 
eigenvector is mostly $\rho$ relaxes 
slowly compared to the other which is predominantly $\mn$. 
This makes it possible to make the dynamics of $\mn$ slaved to $\rho$ and obtain a meaningful 
effective equation for the latter. We will show that the $\rho$ dynamics is equivalent to that of an electrolyte 
or a gravitational system in a passive frictional medium for $G<0$ or $G>0$ 
respectively. 

$G<0$ implies a repulsive interaction between the swimmers; $\rho$ and $-{G}/{2D_r'}$ in this case are analogous to 
the charge density and Ohmic conductivity respectively in an electrolyte. 
Remember that the static structure factor is simply the strength of fluctuations divided by the 
rate at which they decay; for $\rho$ the fluctuations are number conserving so that they 
go as $q^2$ while the relaxation is anomalously fast for wavelengths larger than a screening length 
due to the non local nature of the Coulomb-like interactions. 
To obtain the structure factors requires, of course, that we add phenomenological, gaussian, spatiotemporally
white noise terms, conserving for $\rho$ and nonconserving for $\mP_L$ of strength $\mathcal{P}$ and $\mathcal{N}$, to the
equations of motion. 
The full expressions for the structure factors are
\begin{eqnarray}\label{sqsat}
S_\rho &=& q^2 \frac{2 \mathcal{P}  +   \mathcal{N} (v_0 + \epsilon \rho_0 B)^2/2 D_r ^2 }{-G/2D_r + \left[\frac{v_0(v_0 + \epsilon \rho_0 B)}{3 D_r}  + G D/4 D_r' D_r\right] q^2 + ...} \nonumber \\
S_n &=& \frac{2 B^2 \mathcal{P}/D_r^2   +   \mathcal{N} A^2/9D_r^2 }{-G/2D_r + \left[\frac{v_0(v_0 + \epsilon \rho_0 B)}{3 D_r}  + G D/4 D_r' D_r \right] q^2 + ...}
\end{eqnarray}
neglecting terms that are small when $D_r$ is large. There is screening behaviour where fluctuations 
at length scales larger than 
\begin{eqnarray}\label{qth}
\lambda_{s} = \sqrt{\frac{2D_r[\frac{v_0(v_0 + \epsilon \rho_0 B)}{3 D_r}  + G D/4 D_r' D_r]}{-G}}
\end{eqnarray}
are suppressed in an essentially wavenumber independent way. 
Including terms of higher order in $q$ indicated in the ellipsis in \eqref{sqsat} yields 
a density structure factor with a peak at $q \sim G^{1/4}$, characteristic of 
microphase separation. Fluctuations in $\mn$ are suppressed and $S_{n}$ just decays.

For $G>0$, the mode structure in \eqref{saturatedmodes} 
shows that there is an instability for $G>0$ so that the structure factors in \eqref{sqsat} have 
to be interpreted differently now. For wavenumbers larger than $\lambda_s^{-1}$
the fluctuations die off whereas for wavenumbers smaller than $\lambda_s^{-1}$ the system 
self gravitates and collapses; this is a dissipative jeans instability. $S_n$ 
also shows a peak as density gradients produced by $\rho$ tends to align $\mn$.
These effects are reminiscent of behaviour predicted for a collection of thermophoretic 
colloids \cite{thermophoretic2012}.
 
Increasing $A$ or $B$ and thus slowing down the relaxation of $\mn_L$ takes one out of
the overdamped regime. The general expressions for the eigenfrequencies are 
\begin{eqnarray}\label{modeosc}
-i\omega &=& -{D_r'} - Dq^2  \pm  \sqrt{(D_r^{'2} + G) - \frac{v_0 (v_0 + \epsilon \rho_0 B)
q^2}{3}},
\end{eqnarray}
which can now become oscillatory. In the underdamped limit $\mn$ behaves like a velocity field and 
the systems resembles a system conserving momentum with some subtle differences discussed in 
context of the advective nonlinearities in the Appendix. For $G>0$ and $D_r' \to 0$ we see 
in \eqref{modeosc} that a mode structure similar to the known hydrodynamic Jeans instability is obtained 
with corrections due to disspation. Modes with wavenumber larger than $q^{*}$
\begin{eqnarray}
q^{*} = \sqrt{\frac{3(D_r^{'2}+G)}{v_0 (v_0 + \epsilon \rho_0 B)}}
\end{eqnarray}
are oscillatory and the ones smaller than $q^{*}$ are too `massive' and collapse. Note that the crossover 
wavenumber is given by a competition between the interaction strength $G$ and sound speed square equivalent $v_0 (v_0 + \epsilon \rho_0 B)$ 
for our system. For $(G+D_r^{'2})<0$ 
we now have a electrolyte that shows plasma oscillations with typical frequency $\sqrt{|G|}$ as $D_r \to 0$. 
If in addition $D_r'<0$ it is possible that the oscillations
could become spontaneous. 

The static structure factors on the stable side are
\begin{eqnarray}\label{sqsat}
S_\rho &=& q^2 \frac{\mathcal{N}}{ D_r'+D q^2} + q^2 \frac{4 D_r^2  \mathcal{P}  +   \mathcal{N} (v_0 + \epsilon \rho_0 B)^2}{\left[ \frac{2v_0\left(v_0 + \epsilon \rho_0 B \right)}{3 } q^2 - G\right]( D_r' + D q^2)}, \nonumber \\
S_n &=& \frac{\mathcal{N}}{ D_r'+D q^2} + \frac{ B^2 \mathcal{P}   +   \mathcal{N} A^2/9 }{\left[ \frac{2v_0\left(v_0 + \epsilon \rho_0 B \right)}{3 } q^2 - G\right]}.
\end{eqnarray}
We find a structure factor for $\mP_L$ with correlation length $\sim
(D/D_r')^{1/2}$ indicating strong fluctuations towards aster formation. 
On approaching $D_r' \to 0$, from the stable side the peak values of $S_\rho$ and $S_n$ 
increase as seen in \eqref{sqsat}. For $G>0$ and $D_r<0$ we have a jeans instability with a very 
strongly correlated hydrodynamic field; however extensive study of this interesting regime is beyond the scope 
of our present work.
\end{document}